\documentclass[11pt]{article}
\usepackage{graphicx}
\usepackage{yfonts}
\usepackage{tikz}
\usepackage{amsmath}
\usepackage{amssymb}
\usepackage{amsfonts}
\usepackage{braket}
\usepackage{xcolor}
\usepackage{jheppub}
\usepackage{url}
\usepackage{mathtools}
\usepackage{float}
\usepackage{tensor}
\usepackage{listofitems}
\usepackage{forloop}
\usepackage{xstring,xifthen}
\usepackage{multirow}
\usepackage{array}
\usepackage{tabularray}
\usepackage{longtable}
\numberwithin{equation}{section}
\usetikzlibrary{arrows}
\usetikzlibrary{positioning}

\DeclareMathOperator{\Li2}{\text{Li}_2}
\definecolor{cadetblue}{rgb}{0.37, 0.62, 0.63}
\definecolor{beaver}{rgb}{0.62, 0.51, 0.44}
\let\emptyset\varnothing
\newcommand{\zigzag}[1]{
    \readlist*\argslist{#1,\phantom{},\phantom{},\phantom{},\phantom{},\phantom{},\phantom{},\phantom{},\phantom{},\phantom{},\phantom{},\phantom{},\phantom{}}
    \genfrac{}{}{0pt}{}{\phantom{\argslist[1]} \argslist[2] \phantom{\argslist[3]} \argslist[4] \phantom{\argslist[5]} \argslist[6] \phantom{\argslist[7]} \argslist[8] \phantom{\argslist[9]}}{\argslist[1] \phantom{\argslist[2]} \argslist[3] \phantom{\argslist[4]} \argslist[5] \phantom{\argslist[6]} \argslist[7] \phantom{\argslist[8]} \argslist[9]}
}
\newcommand{\C}[5][\phantom{}]{#1\Bigg[\zigzag{#2}; #3 \hspace{0.2em} \Bigg| \hspace{0.2em} \zigzag{#4}; #5\Bigg]}
\newcommand{\Z}[3][\phantom{}] {
    \StrLen{#3}[\arglen]
    #1\Bigg(\zigzag{#2} \ifthenelse{\arglen > 0}{;}{} #3\Bigg)
}
\newcommand{\col}[2]{\genfrac{}{}{0pt}{}{#1}{#2}}

\newtheorem{conjecture}{Conjecture}

\newtheorem{proposition}{Proposition}
\title{}
\author{}
\date{}
\abstract {
    We propose and study blowup relations obeyed by the partition functions of $5d$ $\mathcal{N}=1$ (quiver) SYM theories with $SU(2)$ gauge group and four flavours. By analyzing the Weyl group action on the sets of blowup
    relations, we identify the integer parameters of a blowup relation with the weights of a corresponding Lie algebra. We also explain how this action of the Weyl group
    follows from the Weyl group symmetry of the partition function. Finally, we use these relations to derive bilinear relations on the $q$-Painlev\'e VI tau functions.
}
\begin{document}
\pagenumbering{arabic}
\title{Blowup relations and $q$-Painlev\'e VI}
\author{Artem Stoyan \\ {\it School of Mathematics and Statistics} \\ {\it The University of Melbourne} \\ {\it Parkville, Victoria 3010, Australia}}
\emailAdd{astoian@student.unimelb.edu.au}
\maketitle
\tableofcontents
\section{Introduction}
Blowup relations obeyed by partition functions of supersymmetric gauge theories were proposed and proved in the case without matter in \cite{NY}, \cite{NY2}. These relations are an effective tool for studying gauge theory partition functions. For example, in the original publications \cite{NY}, \cite{NY2} they were used to extract the prepotential from the partition function. 

We are interested in another application of blowup relations to a subject which can be widely described as Painlev\'e/Gauge or Isomonodromy/CFT correspondence. It was started in \cite{GIL} where it was observed that Painlev\'e VI tau function is given by a series of $c=1$ Virasoro conformal blocks (Kiev formula). Since then a lot of progress has been made along the lines of the original proposal and analogous expansions were obtained for various Painlev\'e and $q$-Painlev\'e tau functions \cite{BLMST}, \cite{BS_2017}. The CFT interpretation of the Kiev formula was given in \cite{ILTe}; the solutions of the $q$-deformed isomonodromic problem were constructed in \cite{JNS}. Later it was realized that blowup relations are an effective tool here too. For example, they have been used to obtain the Kiev formulas \cite{JN} and $c=-2$ analogs of them \cite{BS}.
It was observed in \cite{JN_2018}, \cite{JN} that blowup relations naturally relate two Isomonodromy/CFT correspondences, namely the tau function expansion in terms of $c=1$ conformal blocks \cite{GIL} and the relation of quasiclassical conformal blocks to the Painlev\'e action functional \cite{LLNZ}, \cite{GMS}.

In this paper, we propose and study blowup relations that are relevant for the $q$-Painlev\'e VI equation. In terms of the gauge theory, they involve $5d$ $A_1$, $A_2$ and $A_3$ quiver partition functions. 
These gauge theories are considered on the spacetime $\mathbb{C}\times\mathbb{C}\times S^1$, with the Omega background deformation on the complex planes of exponentiated equivariant parameters $q_1,q_2$.
Some of these partition functions are subject to Higgsing, i.e. tuning of fundamental masses. On the CFT side it translates \cite{AGT} to the $4$, $5$, and $6$ point $q$-deformed conformal blocks with some of the primary fields being degenerate. We start with identifying the relevant blowup relations by a series of numerical experiments. Then we study the Weyl group action on the set of blowup relations. Under this action, the integer parameters of a blowup relation transform as weights of the corresponding Lie algebra.  We identify the representations that appear in this way. We also explain how the Weyl group action on the blowup relations arises from the Weyl group symmetry of the partition functions \cite{MPTY}, \cite{FOS}. Then, we use blowup relations to obtain bilinear relations on the $q$-Painlev\'e VI tau functions along with solutions of $q$-Painlev\'e VI. To do so, we consider the $q_2 \to 1$ limit of partition functions and blowup relations. The remaining equivariant parameter $q_1$ is kept fixed and identified with the $q$-deformation parameter. In that aspect our approach generalizes the methods developed in \cite{JN} to the $5$d case. However, we do not generalize the Hamilton-Jacobi equation to the $q$-deformed setting. Instead, we established the set of blowup relations that allows to derive $q$-Painlev\'e VI bilinear relations directly. The limit of a blowup relation becomes an equality that relates a $q$-Painlev\'e VI tau function to an exponential of the generating function $S$ and its derivatives. It allows us to multiply these equalities in such a way that all the contributions of $S$ cancel. The resulting equalities involve only a solution $y(t)$ of the $q$-Painlev\'e VI equation and the tau functions. Eliminating the solution $y(t)$ from these equalities we derive bilinear relations on the tau functions \cite{JNS}.

\paragraph{Notation.}
We set
    \begin{equation}
        \gamma = q_1^{-1/2}q_2^{-1/2}.
    \end{equation}
We use the following notation for the $q$-number
\begin{equation}
    [u]_q = \frac{1-q^u}{1-q}.
\end{equation}
We also use $q$-Gamma function and $q$-Barnes $G$ function defined by
\begin{equation}
    \Gamma_q(u) = \frac{(q;q)_\infty}{(q^u;q)_\infty}(1-q)^{1-u},
\end{equation}
\begin{equation}
    G_q(u) = \frac{(q^u;q,q)_\infty}{(q;q,q)_\infty} (q;q)_\infty^{u-1} (1-q)^{-(u-1)(u-2)/2}.
\end{equation}
They have the following properties
\begin{equation}
    \Gamma_q(u+1)=[u]_q\Gamma_q(u),
\end{equation}
\begin{equation}
    G_q(u+1)=\Gamma_q(u) G_q(u).
\end{equation}
We also recall the definition of the $\theta$-function, and the Jacobi triple product identity
\begin{equation}
    \theta(z;q) = (z;q)_{\infty} (qz^{-1};q)_\infty = \frac{1}{(q;q)_\infty} \sum_{k\in\mathbb{Z}} (-1)^k q^{k(k-1)/2} z^k.
\end{equation}
Finally, we also introduce the double $q$-Pochhammer symbol
\begin{equation}
    (u;q_1,q_2)_\infty = (\gamma^2 u;q_1^{-1},q_2^{-1})_\infty = \exp\Big(-\sum_{k=1}^{\infty}\frac{u^k}{k(1-q_1^k)(1-q_2^k)}\Big) = \prod_{i,j=0}^{\infty}(1-u q_1^i q_2^j)
\end{equation}
and the elliptic gamma function
\begin{equation}
    \Gamma(u;q_1,q_2) = \frac{(q_1q_2u^{-1};q_1,q_2)_\infty}{(u;q_1,q_2)_\infty},
\end{equation}
defined such that
\begin{equation}
    \Gamma(q_1 z;q_1,q_2) = \theta(z;q_2) \Gamma(z;q_1,q_2), \quad \Gamma(q_2 z;q_1,q_2) = \theta(z,q_1) \Gamma(z;q_1,q_2).
\end{equation}

\paragraph{Outline.}
The main results are presented in Section \ref{section:blowup-relations} (Conjectures \ref{conjecture:A1}, \ref{conjecture:A2}, \ref{conjecture:two-term-A2}, \ref{conjecture:A2-higgsed}, \ref{conjecture:A3}, and \ref{conjecture:A3-higgsed}, which introduce new blowup relations for the $5d$ $N_f=4$ instanton partition functions). Conjecture \ref{conjecture:two-term-A2} involves a new type of blowup relation with two terms on the right-hand side. In Section \ref{section:partition-functions}, we review the Weyl group symmetries of the partition functions, equations (\ref{A1:S1}--\ref{A1:S5}) and (\ref{A2:T1}--\ref{A2:S5}), following \cite{MPTY}, and conclude that the same symmetries act on the blowup relations (Propositions \ref{proposition:A1-relations-symmetries} and \ref{proposition:A2-relations-symmetries}). Using this action, we organize the blowup relations into Weyl group orbits, see (\ref{A1-cardinalities}) and (\ref{A2-character}). 
Taking the $q_1^\sigma \to 0$ limit of the blowup relations we find a formula for the coefficients appearing in the lhs of blowup relations (\ref{f-a1-ans}, \ref{f-a2-ans}).
Using this formula, we classify possible blowup relations and ensure that all relevant relations were found.
Finally, in Section \ref{tau-functions-bilinear-relations}, we apply the blowup relations to derive bilinear relations for the $q$-Painlev\'e VI tau functions, equations (\ref{bilinear1}--\ref{bilinear8}).

\section{Partition functions and their symmetries} \label{section:partition-functions}
In this paper we study $\mathcal{N}=1$ $5$d instanton partition functions of $SU(2)$, $SU(2)\times SU(2)$ and $SU(2)\times SU(2)\times SU(2)$ supersymmetric (quiver) gauge theories with fundamental matter multiplets. The $\Omega$-background is parametrized by $q_1=\exp(\varepsilon_1), q_2=\exp(\varepsilon_2)$. This section is devoted to Nekrasov formulas for these partition functions and the Weyl group symmetry of them.
\subsection{\texorpdfstring{$A_1, N_f=4$ partition function}{A-1-partition-function}}
The anti-fundamental masses are given by $e^{m_1}=q_1^{\theta_1+\theta_\infty}$, $e^{m_2}=q_1^{\theta_1-\theta_\infty}$, the fundamental masses by $e^{m_3}=q_1^{\theta_0+\theta_t}$, $e^{m_4}=q_1^{\theta_0-\theta_t}$ and the Coulomb modulus is $\sigma$. In order to have the Weyl group symmetry we use the topological strings partition function \cite{MPTY} rather than the gauge theory partition function. Compared to the gauge theory partition function the topological strings partition function has a slightly different 1-loop part and there is no tree level contribution \cite{MPTY}. Another difference is the $U(1)$ factor (which we denote $\mathcal{F}^{\text{extra}}$ here) related to the fact that the topological string partition function reproduces the $U(2)$ rather than $SU(2)$ instanton part \cite{HKN}.
So we decompose the $A_1$ partition into the following three factors:
\begin{equation} \label{F-4-full}
    \begin{gathered}
        \Z[\mathcal{F}_4]{q_1^{\theta_0}, q_1^{\theta_t}, q_1^{\sigma}, q_1^{\theta_1}, q_1^{\theta_\infty}}{q_1, q_2; t, 1} =  \\
        = \mathcal{F}_{4}^{\text{extra}}(\theta_t,\theta_1,t;q_1,q_2) \Z[\mathcal{F}^{\text{1-loop}}_4]{q_1^{\theta_0}, q_1^{\theta_t}, q_1^{\sigma}, q_1^{\theta_1}, q_1^{\theta_\infty}}{q_1, q_2} \Z[\mathcal{F}_4^\text{inst}]{q_1^{\theta_0}, q_1^{\theta_t}, q_1^{\sigma}, q_1^{\theta_1}, q_1^{\theta_\infty}}{q_1, q_2; t, 1},
    \end{gathered}
\end{equation}
where
\begin{equation}
    \mathcal{F}_{4}^{\text{extra}}(q_1^{\theta_t},q_1^{\theta_1};q_1,q_2;t,1) = \big(t q_1^{\theta_t};q_1,q_2\big)^{-1}_\infty \big(\gamma^{-2} tq_1^{\theta_1}; q_1, q_2 \big)^{-1}_\infty,
\end{equation}
the perturbative part is 1-loop exact
\begin{equation}
    \begin{gathered}
        \Z[\mathcal{F}^{\text{1-loop}}_4]{q_1^{\theta_0}, q_1^{\theta_t}, q_1^{\sigma}, q_1^{\theta_1}, q_1^{\theta_\infty}}{q_1, q_2} = \\
        = \frac{(q_1^{\sigma};q_1,q_2)_\infty (\gamma^{-2} q_1^{\sigma};q_1,q_2)_\infty}{\prod_{\epsilon,\epsilon'=\pm}(\gamma^{-1}q_1^{(\epsilon \theta_0+\epsilon'\theta_t+\sigma)/2}; q_1,q_2)_\infty (\gamma^{-1}q_1^{(\epsilon'\theta_1+\epsilon \theta_\infty+\sigma)/2}; q_1,q_2)_\infty} = \\
        = \frac{\Gamma(q_1^{-\sigma};q_1,q_2)}{\prod_{\epsilon=\pm} \Gamma(\gamma^{-1}q_1^{(\epsilon \theta_0-\theta_t-\sigma)/2};q_1,q_2) \Gamma(\gamma^{-1}q_1^{(-\theta_1+\epsilon \theta_\infty-\sigma)/2};q_1,q_2)} \times \\
        \times \frac{(q_1^{\sigma};q_1,q_2)_\infty (q_1^{-\sigma};q_1,q_2)_\infty}{\prod_{\epsilon,\epsilon'=\pm}(\gamma^{-1} q_1^{(\epsilon \theta_0-\theta_t+\epsilon'\sigma)/2}; q_1,q_2)_\infty (\gamma^{-1} q_1^{(-\theta_1+\epsilon \theta_\infty+\epsilon'\sigma)/2}; q_1,q_2)_\infty}
    \end{gathered}
\end{equation}
and the instanton part of the partition function with $t$ being the instanton counting parameter. It can be computed by localization \cite{N_2002}
\begin{equation}
    \begin{gathered}
        \Z[\mathcal{F}_4^{\text{inst}}]{q_1^{\theta_0},q_1^{\theta_t},q_1^{\sigma},q_1^{\theta_1},q_1^{\theta_\infty}}{t,1} = \\
        = \sum_{\pmb{\lambda}} \Big(\gamma^{-2}t\Big)^{|\lambda_+|+|\lambda_-|} \prod_{a,b=\pm} \frac{\mathcal{N}_{\emptyset,\lambda_a}(\gamma q_1^{(b\theta_0+a\sigma+\theta_t)/2})\mathcal{N}_{\lambda_a,\emptyset}(\gamma q_1^{(\theta_1-a\sigma+b\theta_\infty)/2})}{\mathcal{N}_{\lambda_a,\lambda_b}(q_1^{\sigma(b-a)/2})},
    \end{gathered}
\end{equation}
where the summation goes over the pairs of Young diagrams $\pmb{\lambda}=(\lambda_+,\lambda_-)$.
$\mathcal{N}_{\lambda,\mu}$ are defined for the pairs $(\lambda,\mu)$ of Young diagrams, $\chi_{(i,j)\in\lambda} = q_1^{i-1}q_2^{j-1}$ is the box content, $l(\lambda)$ is the number of rows in a Young diagram, $|\lambda|$ is the number of boxes and $\lambda'$ denotes the transposed diagram. In this notation $\mathcal{N}_{\lambda,\mu}$ is given by
    \begin{equation}
        \begin{gathered}
            \mathcal{N}_{\lambda,\mu}(v_1/v_2) = \prod_{x\in\lambda} (1-q_1q_2 \chi_x/v_2)\prod_{y\in\mu} (1-v_1/\chi_y) \prod_{\substack{x \in \lambda\\y \in \mu}} \frac{(1-q_1\chi_x/\chi_y)(1-q_2\chi_x/\chi_y)}{(1-\chi_x/\chi_y)(1-q_1q_2\chi_x/\chi_y)} = \\
            = \prod_{(i,j)\in \lambda}\Big(1-\frac{v_1}{v_2}q_1^{i-\mu'_j}q_2^{\lambda_i-j+1}\Big) \prod_{(k,l)\in\mu}\Big(1-\frac{v_1}{v_2}q_1^{\lambda'_l-k+1}q_2^{-\mu_k+l}\Big).
        \end{gathered}
    \end{equation}

As explained in \cite{MPTY} $\mathcal{F}_4$ is invariant under the Weyl group $W(D_5)$ \footnote{To observe that invariance, one needs to re-expand the partition function as a series in a Weyl group-invariant variable.}.
The Weyl group can be defined by the following relations on the generators $\mathfrak{s}_i, i=1..5$
\begin{equation}
    W(D_5) = <\mathfrak{s}_i|i=1..5>/(\mathfrak{s}_i^2=1, \mathfrak{s}_i \mathfrak{s}_j = \mathfrak{s}_j \mathfrak{s}_i \text{ if } C^{(D_5)}_{ij}=0, \mathfrak{s}_i \mathfrak{s}_j \mathfrak{s}_i = \mathfrak{s}_j \mathfrak{s}_i \mathfrak{s}_j \text{ if } C^{(D_5)}_{ij}=-1),
\end{equation}
where the Cartan matrix is
\begin{equation} \label{D5-Cartan-matrix}
    C^{(D_5)}=
    \begin{pmatrix}
        2 & -1 & 0 & 0 & 0 \\
        -1 & 2 & -1 & 0 & 0 \\
        0 & -1 & 2 & -1 & -1 \\
        0 & 0 & -1 & 2 & 0 \\
        0 & 0 & -1 & 0 & 2
    \end{pmatrix}
\end{equation}
We will also need an extension of the Weyl group $\widetilde{W}(D_5)$ by the diagram automorphism $\sigma_S$.
The action of the Weyl group $\mathfrak{s}_i \mapsto S_i$ on the partition functions is given by (see, e.g. \cite{TM})
\begin{align} 
    \label{A1:S1}
    &S_1: (q_1^{\theta_0},q_1^{\theta_t},q_1^{\sigma},q_1^{\theta_1},q_1^{\theta_\infty},t) \mapsto (q_1^{-\theta_0},q_1^{\theta_t},q_1^{\sigma},q_1^{\theta_1},q_1^{\theta_\infty},t), \\
    \label{A1:S2}
    &S_2: (q_1^{\theta_0},q_1^{\theta_t},q_1^{\sigma},q_1^{\theta_1},q_1^{\theta_\infty},t) \mapsto \\
        & \mapsto (t^{-1}q_1^{(\theta_0-\theta_t-\theta_1-\theta_\infty)/2},q_1^{\theta_t},t q_1^{\sigma+(\theta_0+\theta_t+\theta_1+\theta_\infty)/2},q_1^{\theta_1},t^{-1}q_1^{(\theta_\infty-\theta_0-\theta_t-\theta_1)/2}, q_1^{-(\theta_0+\theta_t+\theta_1+\theta_\infty)/2}), \\
    \label{A1:S3}
    &S_3: (q_1^{\theta_0},q_1^{\theta_t},q_1^{\sigma},q_1^{\theta_1},q_1^{\theta_\infty},t) \mapsto (q_1^{\theta_0},tq_1^{\theta_t},t q_1^{\sigma},tq_1^{\theta_1},q_1^{\theta_\infty},t^{-1}), \\
    \label{A1:S4}
    &S_4: (q_1^{\theta_0},q_1^{\theta_t},q_1^{\sigma},q_1^{\theta_1},q_1^{\theta_\infty},t) \mapsto (q_1^{\theta_0},q_1^{-\theta_t},q_1^{\sigma},q_1^{\theta_1},q_1^{\theta_\infty},tq_1^{\theta_t}), \\
    \label{A1:S5}
    &S_5: (q_1^{\theta_0},q_1^{\theta_t},q_1^{\sigma},q_1^{\theta_1},q_1^{\theta_\infty},t) \mapsto (q_1^{\theta_0},q_1^{\theta_t},q_1^{\sigma},q_1^{-\theta_1},q_1^{\theta_\infty},tq_1^{\theta_1}).
\end{align}
The starting point for the Weyl group symmetry is an observation that an $SU(2)$ theory with $N_f\leq7$ flavors has $SO(2N_f)$ symmetry corresponding to the permutation of masses. It was proposed by Seiberg \cite{S_1996} that these theories have a UV fixed point at which the global symmetry is enhanced to $E_{N_f+1}$.
In \cite{MPTY} the symmetry enhancement was checked at the level of partition functions. It was argued that the enhancement is due to the fiber-base duality \cite{KMV}, \cite{AHK}, \cite{BPTY}. In terms of $(p,q)$-brane realization, this symmetry corresponds to S-duality exchanging D5-brane and NS5-brane. 
The calculation of refined topological strings amplitudes using the topological
vertex requires the choice of a preferred direction of the toric diagram \cite{IKV}. It has been conjectured, and then
proved in \cite{FOS}, that the amplitude does not depend on this choice. The change of preferred
direction is related to the fiber-base duality of the Calabi-Yau, and to the S-duality of the corresponding gauge theory.
The Weyl group symmetry of the partition function gives rise to the action on blowup relations which we discuss in the next section.
\subsection{\texorpdfstring{$A_2, N_f=4$ partition function}{A-2-partition-function}}
Now we consider $A_2$ quiver partition function. There are two instanton counting parameters denoted by $t$ and $y^{-1}$, two Coulomb moduli $\sigma_1$, $\sigma_2$ and the masses parametrized by $\theta_0,\theta_t,\theta_1,\theta_y,\theta_\infty$,
Similarly to the $A_1$ case (see the previous section) we have the decomposition
\begin{equation} \label{F-5-full}
    \mathcal{F}_5 = \mathcal{F}^{\text{extra}}_5 \mathcal{F}^{\text{1-loop}}_5 \mathcal{F}_5^{\text{inst}},
\end{equation}
where
\begin{equation}
    \begin{gathered}
        \Z[\mathcal{F}^{\text{extra}}_5]{q_1^{\theta_0}, q_1^{\theta_t}, q_1^{\sigma_1}, q_1^{\theta_1}, q_1^{\sigma_2}, q_1^{\theta_y}, q_1^{\theta_\infty}}{q_1, q_2;t,1,y} = \\
        = \frac{(\gamma^{-1} t q_1^{(\sigma_1+\theta_1-\sigma_2)/2};q_1,q_2)_\infty (\gamma^{-1} t q_1^{\theta_t+(\sigma_1+\theta_1-\sigma_2)/2};q_1,q_2)_\infty}{(t q_1^{\theta_t};q_1,q_2)_\infty (\gamma^{-2} t q_1^{\theta_1};q_1,q_2)_\infty (ty^{-1} q_1^{\theta_1+\theta_t};q_1,q_2)_\infty (\gamma^{-2} ty^{-1} q_1^{\theta_1+\theta_y};q_1,q_2)_\infty} \times \\
        \times \frac{(\gamma^{-1} y^{-1} q_1^{(-\sigma_1+\theta_1+\sigma_2)/2};q_1,q_2)_\infty (\gamma^{-1} y^{-1} q_1^{\theta_y+(-\sigma_1+\theta_1+\sigma_2)/2};q_1,q_2)_\infty}{(y^{-1} q_1^{\theta_1};q_1,q_2)_\infty (\gamma^{-2} y^{-1} q_1^{\theta_y};q_1,q_2)_\infty},
    \end{gathered}
\end{equation}
\begin{equation}
    \begin{gathered}
        \Z[\mathcal{F}^{\text{1-loop}}_5]{q_1^{\theta_0}, q_1^{\theta_t}, q_1^{\sigma_1}, q_1^{\theta_1}, q_1^{\sigma_2}, q_1^{\theta_y}, q_1^{\theta_\infty}}{q_1, q_2} = \\
        = \frac{\prod_{\epsilon=\pm}(\gamma^{-1-\epsilon}q_1^{\sigma_1};q_1,q_2)_\infty (\gamma^{-1-\epsilon}q_1^{\sigma_2};q_1,q_2)_\infty}{\prod_{\epsilon=\pm}(\gamma^{-1} q_1^{(\sigma_1+\epsilon\theta_1+\sigma_2)/2}; q_1,q_2)_\infty} \times \\
        \times \Big(\prod_{\epsilon,\epsilon'=\pm}(\gamma^{-1} q_1^{(\epsilon \theta_0+\epsilon'\theta_t+\sigma_1)/2}; q_1,q_2)_\infty (\gamma^{-1} q_1^{(\sigma_2+\epsilon \theta_y+\epsilon'\theta_\infty)/2}; q_1,q_2)_\infty\Big)^{-1} = \\
        = (\gamma^{-1} q_1^{(\sigma_1-\theta_1-\sigma_2)/2};q_1,q_2)_\infty (\gamma^{-1} q_1^{(\sigma_2-\theta_1-\sigma_1)/2};q_1,q_2)_\infty \times \\
        \times \frac{\Gamma(q_1^{-\sigma_1};q_1,q_2)\Gamma(q_1^{-\sigma_2};q_1,q_2)}{\Gamma(\gamma^{-1} q_1^{(-\sigma_1-\theta_1-\sigma_2)/2};q_1,q_2)} \times \\
        \times \Big(\prod_{\epsilon=\pm}\Gamma(\gamma^{-1} q_1^{(\epsilon \theta_0-\theta_t-\sigma_1)/2};q_1,q_2) \Gamma(\gamma^{-1} q_1^{(-\theta_y+\epsilon \theta_\infty-\sigma_2)/2};q_1,q_2)\Big)^{-1} \times \\
        \times \frac{\prod_{\epsilon=\pm}(q_1^{\epsilon \sigma_1};q_1,q_2)_\infty (q_1^{\epsilon \sigma_2};q_1,q_2)_\infty}{\prod_{\epsilon,\epsilon'=\pm} (\gamma^{-1} q_1^{(\epsilon'\sigma_1-\theta_1+\epsilon\sigma_2)/2}; q_1,q_2)_\infty} \times \\
        \times \Big(\prod_{\epsilon,\epsilon'=\pm} (\gamma^{-1} q_1^{(\epsilon \theta_0-\theta_t+\epsilon'\sigma_1)/2}; q_1,q_2)_\infty (\gamma^{-1} q_1^{(\epsilon \sigma_2-\theta_y+\epsilon'\theta_\infty)/2}; q_1,q_2)_\infty \Big)^{-1}
    \end{gathered}
\end{equation}
and
\begin{equation} \label{partition_function}
    \begin{gathered}
        \Z[\mathcal{F}_5^{\text{inst}}]{q_1^{\theta_0}, q_1^{\theta_t}, q_1^{\sigma_1}, q_1^{\theta_1}, q_1^{\sigma_2}, q_1^{\theta_y}, q_1^{\theta_\infty}}{q_1,q_2; t,1,y} = \\
        = \sum_{\pmb{\lambda},\pmb{\mu}} \Big(\gamma^{-2}t\Big)^{|\lambda_+|+|\lambda_-|} \Big(\gamma^{-2}y^{-1}\Big)^{|\mu_+|+|\mu_-|} \times \\
        \times \prod_{a,b=\pm} \frac{\mathcal{N}_{\emptyset,\lambda_a}(\gamma q_1^{(b\theta_0+\theta_t+a\sigma_1)/2})\mathcal{N}_{\lambda_a,\mu_b}(\gamma q_1^{(-a\sigma_1+\theta_1+b\sigma_2)/2})\mathcal{N}_{\mu_b,\emptyset}(\gamma q_1^{(-b\sigma_2+\theta_y-a\theta_\infty)/2})}{\mathcal{N}_{\lambda_a,\lambda_b}(q_1^{\sigma_1(b-a)/2})\mathcal{N}_{\mu_a,\mu_b}(q_1^{\sigma_2(b-a)/2})}.
    \end{gathered}
\end{equation}
Now the instanton summation goes over two pairs of Young diagrams $\pmb{\lambda} = (\lambda_+,\lambda_-)$ and $\pmb{\mu}=(\mu_+,\mu_-)$.

As in the $A_1$ case, $\mathcal{F}^{\text{extra}}_5$ is related to partial decoupling of the unitary group symmetry. Although we do not give a physical explanation here we explain how it was computed. Our computation can be summarized in the following three steps.
First, we used the result of \cite{FOS} to construct $SU(2)\times SU(2)$ and $SU(3)$ partition functions that are equivalent under the change of the parameters corresponding to a change of the preferred direction. This equivalence follows from an algebraic construction of the topological vertex \cite{AFS}, and an equality of the corresponding matrix elements \cite{FOS}.
Second, we took a Weyl group invariant $SU(3)$ topological strings partition function from \cite{MPTY} and computed the ratio of the invariant $SU(3)$ function and the algebraic counterpart - the product of matrix elements as in step $1$. In this way, we obtained the correct normalization for the invariant quiver partition function (after applying the change of the preferred direction again) 
Finally, we multiplied the result by a Weyl group invariant \footnote{See the Weyl group action below.} expression to make our formula simpler. This factor is
\begin{equation}
    \begin{gathered}
        (\gamma^{-1}q_1^{(\sigma_1-\theta_1-\sigma_2)/2};q_1,q_2)_\infty (\gamma^{-1}q_1^{(\sigma_1+\theta_1-\sigma_2)/2};q_1,q_2)_\infty (y\gamma^{-1}q_1^{(\sigma_1-\theta_1-\sigma_2)/2};q_1,q_2)_\infty \times \\
        \times (y\gamma^{-1}q_1^{-\theta_y+(\sigma_1-\theta_1-\sigma_2)/2};q_1,q_2)_\infty (t \gamma^{-1}q_1^{(\sigma_1+\theta_1-\sigma_2)/2};q_1,q_2)_\infty (t \gamma^{-1}q_1^{\theta_t+(\sigma_1+\theta_1-\sigma_2)/2};q_1,q_2)_\infty.
    \end{gathered}
\end{equation}
Another reason for introducing this factor is to cancel the double Pochhammer symbols having $y$ in the argument. Since the instanton part is expanded in the powers of $y^{-1}$ these factors would spoil the convergence.
The full partition function is invariant under the action of the Weyl group $W(A_1 \times A_1\times A_5)$ \cite{MPTY}. Compared to $W(A_5)$ there are two additional generators $\mathfrak{t}_1$ and $\mathfrak{t}_2$ which commute with $W(A_5)$ and each other. The action of the generators $\mathfrak{s}_i \mapsto S_i, \mathfrak{t}_i \mapsto T_i$ is listed below.
\begin{align} 
    \label{A2:T1}
    &T_1: (q_1^{\theta_0},q_1^{\theta_t},q_1^{\sigma_1}, q_1^{\theta_1},q_1^{\sigma_2}, q_1^{\theta_y},q_1^{\theta_\infty},t,y) \mapsto (q_1^{-\theta_0},q_1^{\theta_t},q_1^{\sigma_1}, q_1^{\theta_1},q_1^{\sigma_2},q_1^{\theta_y},q_1^{\theta_\infty},t,y), \\
    \label{A2:T2}
    &T_2: (q_1^{\theta_0},q_1^{\theta_t},q_1^{\sigma_1}, q_1^{\theta_1},q_1^{\sigma_2}, q_1^{\theta_y},q_1^{\theta_\infty},t,y) \mapsto (q_1^{\theta_0},q_1^{\theta_t},q_1^{\sigma_1}, q_1^{\theta_1},q_1^{\sigma_2},q_1^{\theta_y},q_1^{-\theta_\infty},t,y), \\
    \label{A2:S1}
    &S_1: (q_1^{\theta_0},q_1^{\theta_t},q_1^{\sigma_1}, q_1^{\theta_1},q_1^{\sigma_2}, q_1^{\theta_y},q_1^{\theta_\infty},t,y) \mapsto (q_1^{\theta_0},q_1^{-\theta_t},q_1^{\sigma_1},q_1^{\theta_1},q_1^{\sigma_2},q_1^{\theta_y},q_1^{\theta_\infty},tq_1^{\theta_t},y), \\
    \label{A2:S2}
    &S_2: (q_1^{\theta_0},q_1^{\theta_t},q_1^{\sigma_1}, q_1^{\theta_1},q_1^{\sigma_2}, q_1^{\theta_y},q_1^{\theta_\infty},t,y) \mapsto (q_1^{\theta_0},tq_1^{\theta_t},t q_1^{\sigma_1},tq_1^{\theta_1},q_1^{\sigma_2},q_1^{\theta_y},q_1^{\theta_\infty},t^{-1},y), \\
    \label{A2:S3}
    &S_3: (q_1^{\theta_0},q_1^{\theta_t},q_1^{\sigma_1}, q_1^{\theta_1},q_1^{\sigma_2}, q_1^{\theta_y},q_1^{\theta_\infty},t,y) \mapsto (q_1^{\theta_0},q_1^{\theta_t},q_1^{\sigma_1},q_1^{-\theta_1},q_1^{\sigma_2},q_1^{\theta_y},q_1^{\theta_\infty},t q_1^{\theta_1},yq_1^{-\theta_1}), \\
    \label{A2:S4}
    &S_4: (q_1^{\theta_0},q_1^{\theta_t},q_1^{\sigma_1}, q_1^{\theta_1},q_1^{\sigma_2}, q_1^{\theta_y},q_1^{\theta_\infty},t,y) \mapsto (q_1^{\theta_0},q_1^{\theta_t},q_1^{\sigma_1},y^{-1}q_1^{\theta_1},y^{-1}q_1^{\sigma_2},y^{-1}q_1^{\theta_y},q_1^{\theta_\infty},t,y^{-1}), \\
    \label{A2:S5}
    &S_5: (q_1^{\theta_0},q_1^{\theta_t},q_1^{\sigma_1}, q_1^{\theta_1},q_1^{\sigma_2}, q_1^{\theta_y},q_1^{\theta_\infty},t,y) \mapsto (q_1^{\theta_0},q_1^{\theta_t},q_1^{\sigma_1},q_1^{\theta_1},q_1^{\sigma_2},q_1^{-\theta_y},q_1^{\theta_\infty},t,yq_1^{-\theta_y}).
\end{align}
As in the $A_1$ case this action gives rise to the action on blowup relations which we discuss below.
\subsection{\texorpdfstring{$A_3,N_f=4$ partition function}{A-3-partition-function}}
We will also need the instanton part of the $A_3$ partition function
\begin{equation}
    \begin{gathered}
        \Z[\mathcal{F}_6^\text{inst}]{q_1^{\theta_0}, q_1^{\theta_x}, q_1^{\sigma_0}, q_1^{\theta_t}, q_1^{\sigma_1}, q_1^{\theta_1}, q_1^{\sigma_2}, q_1^{\theta_y}, q_1^{\theta_\infty}}{q_1, q_2; x, t, 1, y} = \\
        = \sum_{\pmb{\lambda},\pmb{\mu},\pmb{\nu}} \Big(\gamma^{-2}xt^{-1}\Big)^{|\lambda_+|+|\lambda_-|} \Big(\gamma^{-2}t\Big)^{|\mu_+|+|\mu_-|} \Big(\gamma^{-2}y^{-1}\Big)^{|\nu_+|+|\nu_-|} \times \\
        \times \prod_{a,b=\pm} \frac{\mathcal{N}_{\emptyset,\lambda_a}(\gamma q_1^{(-b\theta_0+\theta_x+a\sigma_0)/2})\mathcal{N}_{\lambda_a,\mu_b}(\gamma q_1^{(-a\sigma_0+\theta_t+b\sigma_1)/2})}{\mathcal{N}_{\lambda_a,\lambda_b}(q_1^{\sigma_0(b-a)/2})\mathcal{N}_{\mu_a,\mu_b}(q_1^{\sigma_1(b-a)/2})\mathcal{N}_{\nu_a,\nu_b}(q_1^{\sigma_2(b-a)/2})} \times \\
        \times \prod_{a,b=\pm} \mathcal{N}_{\mu_b,\nu_a}(\gamma q_1^{(-b\sigma_1+\theta_1+a\sigma_2)/2})\mathcal{N}_{\nu_a,\emptyset}(\gamma q_1^{(-a\sigma_2+\theta_y+b\theta_\infty)/2}).
    \end{gathered}
\end{equation}
\section{Blowup relations} \label{section:blowup-relations}
\subsection{\texorpdfstring{$A_1$ relations}{A1}}
We start with $A_1$ blowup relations. We present the following conjecture based on a series of numerical experiments, i.e. order by order checks of the expansion in the powers of instanton counting parameters.
\begin{conjecture} \label{conjecture:A1}
We have found $83$ relations in total. Each relation (\ref{blowup4}) is described by: the value of $\nu=0$ (integer relations) or $\nu=1/2$ (half-integer relations) and a tuple $(j_1,j_2,j_3,j_4,d) \in \mathbb{Z}^5$ giving the shifts of the mass parameters and the instanton counting parameter. 
Each relation contains a rational function $f$ of $q_1^{\pmb{\theta}/2}$, $q_1^{d/2}$, $q_1^{1/2}$, $q_2^{1/2}$ and $t$ appearing on the lhs. The coefficients $\mathcal{C}_4$ can be computed as ratios of the 1-loop contributions to the gauge theory partition function, see (\ref{C_4}) for the formula.
\begin{equation}
    \begin{gathered} \label{blowup4}
        f(\pmb{\theta},t|\pmb{j},d)\times\Z[\mathcal{F}_4^\text{inst}]{q_1^{\theta_0}, q_1^{\theta_t}, q_1^{\sigma}, q_1^{\theta_1}, q_1^{\theta_\infty}}{q_1, q_2; t, 1} = \sum_{n\in\mathbb{Z}+\nu} \C[\mathcal{C}_4]{\theta_0, \theta_t, \sigma, \theta_1, \theta_\infty}{t}{j_1, j_2, n, j_3, j_4}{d} \times \\
        \times \Z[\mathcal{F}_4^\text{inst}]{q_1^{\theta_0+j_1}, q_1^{\theta_t+j_2}, q_1^{\sigma+2n}, q_1^{\theta_1+j_3}, q_1^{\theta_\infty+j_4}}{q_1, q_1^{-1}q_2; t q_1^d, 1} \times \\
        \times \Z[\mathcal{F}_4^\text{inst}]{q_1^{\theta_0}q_2^{j_1}, q_1^{\theta_t}q_2^{j_2}, q_1^{\sigma}q_2^{2n}, q_1^{\theta_1}q_2^{j_3}, q_1^{\theta_\infty}q_2^{j_4}}{q_1q_2^{-1}, q_2; t q_2^d, 1}.
    \end{gathered}
\end{equation}
The value of $\nu$ is determined by the parity of $j_1+j_2$ (which is the same as the parity of $j_3+j_4$).
\begin{equation} \label{a1-parity}
    \nu = \frac{1}{2} \Big(j_1+j_2+1 \mod 2\Big) = \frac{1}{2}\Big(j_3+j_4+1 \mod 2\Big)
\end{equation}
For the list of all relations see Table \ref{A1-table}. This table contains the corresponding function $f$ for each of the relations.
\end{conjecture}
We now observe that the given set of blowup relations is invariant under the action of $W(D_5)$. The action is given in the following proposition. Although the value of $\nu$ is determined by the value of $j_1+j_2$ we state the transformation of $n$ in the formula below as it follows from the transformation of $q_1^\sigma$ given in (\ref{A1:S1} -- \ref{A1:S5}). This transformation of $n$ should be used in the formula (\ref{A1:f_action}). This transformation agrees with $\nu = n \mod 2$.
\begin{proposition} \label{proposition:A1-relations-symmetries}
    The extended Weyl group $\widetilde{W}(D_5)$ acts on the set of blowup relations. The generators act on the integer parameters $(j_1, j_2, n, j_3, j_4, d)$ by $\mathfrak{s}_i \mapsto s_i$ where
    \begin{equation} \label{A1:s1}
        s_1: (j_1,j_2,n,j_3,j_4,d) \mapsto (-j_1,j_2,n,j_3,j_4,d),
    \end{equation}
    \begin{equation} \label{A1:s2}
        \begin{gathered}
            s_2: (j_1,j_2,n,j_3,j_4,d) \mapsto \Big(\frac{1}{2}\Big(j_1-j_2-j_3-j_4\Big)-d,j_2,n+\frac{d}{2}+\frac{1}{4}\Big(j_1+j_2+j_3+j_4\Big), \\
            j_3,-\frac{1}{2}\Big(j_1+j_2+j_3-j_4\Big)-d,-\frac{1}{2}\Big(j_1+j_2+j_3+j_4\Big)\Big),
        \end{gathered}
    \end{equation}
    \begin{equation} \label{A1:s3}
        s_3: (j_1,j_2,n,j_3,j_4,d) \mapsto (j_1,j_2 + d,n+\frac{d}{2}, j_3 + d,j_4,-d),
    \end{equation}
    \begin{equation} \label{A1:s4}
        s_4: (j_1,j_2,n,j_3,j_4,d) \mapsto (j_1,-j_2,n,j_3,j_4,d+j_2),
    \end{equation}
    \begin{equation} \label{A1:s5}
        s_5: (j_1,j_2,n,j_3,j_4,d) \mapsto (j_1,j_2,n,-j_3,j_4,d+j_3),
    \end{equation}
    \begin{equation} \label{A1:sigma_S}
        \sigma_S: (j_1,j_2,j_3,j_4,d) \mapsto (j_1,j_3,j_2,j_4,d).
    \end{equation}
$f$-factors in the lhs of (\ref{blowup4}) transform according to
    \begin{equation} \label{A1:f_action}
        \begin{gathered}
            f(\pmb{\theta},t|s_i(\pmb{j},d)) = S_i \Bigg(f(\pmb{\theta},t|\pmb{j},d) \C[\mathcal{A}_4]{\theta_0,\theta_t,\sigma,\theta_1,\theta_\infty}{t}{j_1,j_2,n,j_3,j_4}{d}^{-1}\Bigg) \times \\
            \times s_i \Bigg(\C[\mathcal{A}_4]{\theta_0,\theta_t,\sigma,\theta_1,\theta_\infty}{t}{j_1,j_2,n,j_3,j_4}{d}\Bigg),
        \end{gathered}
    \end{equation}
where
    $\mathcal{A}_4$ is defined in (\ref{A_4}) and $S_i$  act by (\ref{A1:S1}--\ref{A1:S5}).
\end{proposition}
The action of the non-extended Weyl group on the set of blowup relations comes from the symmetries of the partition function given by (\ref{A1:S1}--\ref{A1:S5}). This observation immediately leads to the formula (\ref{A1:f_action}) for the transformation of $f$ coefficients.

To illustrate the action of the Weyl group on the set of blowup relations we pick a blowup relation from Table \ref{A1-table}:
\begin{equation}
    f = \frac{1}{(1-tq_1^{\theta_1})}, \quad (j_1,j_2,n,j_3,j_4,d) = (0,1,0,0,-1,-1)
\end{equation}
Note that we set $n=0$ because $j_1+j_2+1 \mod 2 = 0$ so we have an integer $(\nu = 0)$ relation here. We could have set $n$ to be any integer, but we choose $n=0$ to simplify the computation. We are going to apply $s_2$ to this relation.
First of all we apply (\ref{A1:s2}) and obtain
\begin{equation}
    s_2: (0,1,0,0,-1,-1) \mapsto (1,1,-\frac{1}{2},0,0,0)
\end{equation}
In particular, we see that $s_2$ applied to that integer relation gives a half-integer relation since we got $n=-1/2$. To compute the transformed $f$ we use the formula (\ref{A1:f_action}). It reads
\begin{equation}
    \begin{gathered}
        S_2: \frac{1}{(1-tq_1^{\theta_1})} \mapsto S_2 \Bigg(\frac{1}{(1-tq_1^{\theta_1})} \C[\mathcal{A}_4]{\theta_0,\theta_t,\sigma,\theta_1,\theta_\infty}{t}{0,1,0,0,-1}{-1}^{-1}\Bigg) \times \\
        \times \C[\mathcal{A}_4]{\theta_0,\theta_t,\sigma,\theta_1,\theta_\infty}{t}{1,1,-\frac{1}{2},0,0}{0} = \\
                = \frac{1}{(1-q_1^{(-\theta_0-\theta_t+\theta_1-\theta_\infty)/2})} \times \\
                \times \C[\mathcal{A}_4]{\frac{\theta_0-\theta_t-\theta_1-\theta_\infty}{2} - \frac{\log t}{\log q_1},\theta_t,\sigma+\sum \frac{\theta_i}{2},\theta_1,\frac{\theta_\infty-\theta_0-\theta_t-\theta_1}{2} - \frac{\log t}{\log q_1}}{q_1^{-\sum \theta_i/2}}{0,1,0,0,-1}{-1}^{-1} \times \\
                \times \C[\mathcal{A}_4]{\theta_0,\theta_t,\sigma,\theta_1,\theta_\infty}{t}{1,1,-\frac{1}{2},0,0}{0} = \\
                = -\frac{(q_1 q_2 t q_1^{\theta_t} q_1^{\theta_1})^{1/4}}{1-tq_1^{\theta_t}} = \\
                = \frac{1}{1-tq_1^{\theta_t}}\Bigg(\C[\mathcal{C}_4]{\theta_0,\theta_t,\sigma,\theta_1,\theta_\infty}{t}{1,1,\frac{1}{2},0,0}{0} + \C[\mathcal{C}_4]{\theta_0,\theta_t,\sigma,\theta_1,\theta_\infty}{t}{1,1,-\frac{1}{2},0,0}{0}\Bigg),
    \end{gathered}
\end{equation}
which agrees with the value given in Table \ref{A1-table}.

Using the Cartan matrix (\ref{D5-Cartan-matrix}) and transformations (\ref{A1:s1}--\ref{A1:s5}) we compute the simple roots $\alpha_i$ and fundamental weights $\omega_i$ of $D_5$. The simple roots and the Killing form satisfy\footnote{Since the Cartan matrix is symmetric and all simple roots are normalized to have squared length $2$.}
\begin{equation} \label{roots-scalar-product}
    (\alpha_i,\alpha_j) = \alpha_i K \alpha_j^T = C^{(D_5)}_{ij}
\end{equation}
and
\begin{equation} \label{root-reflection}
    s_i \alpha_j = \alpha_j - (\alpha_i,\alpha_j)\alpha_i
\end{equation}
One can solve the system of equations (\ref{roots-scalar-product}, \ref{root-reflection}) for both the simple roots and the Killing form in the coordinates $(j_1,j_2,j_3,j_4,d)$. The result reads
\begin{align} \label{D4:roots}
    \begin{aligned}
        \alpha_1 &= (2,0,0,0,0), \\
        \alpha_2 &= (-1,0,0,-1,-1), \\
        \alpha_3 &= (0,-1,-1,0,2), \\
        \alpha_4 &= (0,2,0,0,-1), \\
        \alpha_5 &= (0,0,2,0,-1),
    \end{aligned}
\end{align}
and
\begin{equation}
    K^{(D_5)} = 
    \begin{pmatrix}
        \frac{1}{2} & 0 & 0 & 0 & 0 \\
        0 & \frac{3}{4} & \frac{1}{4} & 0 & \frac{1}{2} \\
        0 & \frac{1}{4} & \frac{3}{4} & 0 & \frac{1}{2} \\
        0 & 0 & 0 & \frac{1}{2} & 0 \\
        0 & \frac{1}{2} & \frac{1}{2} & 0 & 1
    \end{pmatrix}
\end{equation}
Now we write
\begin{equation}
    (\omega_i, \alpha_j) = \delta_{ij}
\end{equation}
and use the Killing form and simple roots to find the fundamental weights in the same coordinates:
\begin{equation} \label{D4:weights}
    \begin{aligned}
        \omega_1 &= (1,0,0,-1,0), \\
        \omega_2 &= (0,0,0,-2,0), \\
        \omega_3 &= (0,0,0,-2,1), \\
        \omega_4 &= (0,1,0,-1,0), \\
        \omega_5 &= (0,0,1,-1,0).
    \end{aligned}
\end{equation}
As a consistency check, we verified that (\ref{D4:roots}, \ref{D4:weights}) agrees with
\begin{equation}
    (\sum_{i=1}^5 \mathbb{Z} \omega_i)/(\sum_{i=1}^5 \mathbb{Z} \alpha_i) = \mathbb{Z}/4\mathbb{Z}.
\end{equation}
We define the invariant quadratic form on the variables $(j_1,j_2,j_3,j_4,d)$ by
\begin{equation}
    Q^{(D^5)}(\pmb{j},d) = (j_1,j_2,j_3,j_4,d)K(j_1,j_2,j_3,j_4,d)^{T}
\end{equation}
It reads
\begin{equation} \label{a1-quadratic-form}
    Q^{(D^5)}(\pmb{j},d) = \frac{j_1^2+j_2^2+j_3^2+j_4^2}{2} + \Big(d+\frac{j_2+j_3}{2}\Big)^2
\end{equation}
The orbits of the extended Weyl group action on the blowup relations have cardinalities
\begin{equation} \label{A1-cardinalities}
    83 = 40+32+10+1.
\end{equation}
The numbers in (\ref{A1-cardinalities}) are dimensions of the fundamental representations of $SO(10)$.
The component consisting of 40 elements can be identified with the root system $D_5$, those tuples of $(\pmb{j},d)$ are weights of the adjoint representation. Its highest weight is $\omega_2$. The orbit of size $32=16+16$ consists of the weights of two spin representations with highest weights $\omega_4$ and $\omega_5$. They are in the same orbit due to the presence of the diagram automorphism $\sigma_S$ among the symmetry generators. The orbit of size $10$ consists of the weights of the defining representation of $SO(10)$. Since we know the fundamental weights in the given coordinates (\ref{D4:weights}), we can directly identify each blowup relation in Table \ref{A1-table} with a weight of an $SO(10)$ representation. This identification refines the formula (\ref{A1-cardinalities}). 

So far we have seen all the fundamental weights but $\omega_3$. It turns out that there are some blowup relations which are related to this representation, but they are more complicated than the ones we have seen above. Here is an example of such relation
\begin{equation} \label{blowup4-2-term}
    \begin{gathered}
        \sum_{j_4=\pm1}j_4q_1^{j_4\theta_\infty/2} \Bigg(\C[\mathcal{C}_4]{\theta_0,\theta_t, \sigma, \theta_1, \theta_\infty}{t}{j_1,j_2,1/2,j_3,j_4}{d}+\C[\mathcal{C}_4]{\theta_0,\theta_t, \sigma, \theta_1, \theta_\infty}{t}{j_1,j_2,-1/2,j_3,j_4}{d}\Bigg) \times \\
        \times \Z[\mathcal{F}_4^\text{inst}]{q_1^{\theta_0}, q_1^{\theta_t}, q_1^{\sigma}, q_1^{\theta_1}, q_1^{\theta_\infty}}{q_1, q_2; t, 1} = \\
        = \sum_{j_4=\pm} j_4q_1^{j_4\theta_\infty/2} \sum_{n\in\mathbb{Z}+1/2} \C[\mathcal{C}_4]{\theta_0, \theta_t, \sigma, \theta_1, \theta_\infty}{t}{j_1, j_2, n, j_3, j_4}{d} \times \\
        \times \Z[\mathcal{F}_4^\text{inst}]{q_1^{\theta_0+j_1}, q_1^{\theta_t+j_2}, q_1^{\sigma+2n}, q_1^{\theta_1+j_3}, q_1^{\theta_\infty+j_4}}{q_1, q_1^{-1}q_2; t q_1^d, 1} \times \\
        \times \Z[\mathcal{F}_4^\text{inst}]{q_1^{\theta_0}q_2^{j_1}, q_1^{\theta_t}q_2^{j_2}, q_1^{\sigma}q_2^{2n}, q_1^{\theta_1}q_2^{j_3}, q_1^{\theta_\infty}q_2^{j_4}}{q_1q_2^{-1}, q_2; t q_2^d, 1},  
    \end{gathered}
\end{equation}
where
\begin{equation}
    (j_1,j_2,j_3,j_4,d) = (-1, -1, 1, \pm 1, -1).
\end{equation}
The novelty here is that the rhs now contains two terms corresponding to $j_4=\pm 1$. This form of a relation is natural to 
expect if we start with an $A_2$ relation (see the next section) and take the limit $y\to \infty$. Given the Weyl group action 
it is possible to generate more general relations starting from the one obtained by this procedure. In particular, one of these 
relations involves the fundamental weight $\omega_3$. We do not study these relations here and hope to address this problem 
elsewhere.

There is a formula allowing to compute $f(\pmb{\theta},t|\pmb{j},d)$.

\begin{proposition} \label{proposition:f-a1-ans}
    Assume that Conjecture \ref{conjecture:A1} holds.
    Then the factors $f(\pmb{\theta},t|\pmb{j},d)$ in (\ref{blowup4}) are given by
    \begin{equation} \label{f-a1-ans}
        f(\pmb{\theta},t|\pmb{j},d) = \lim\limits_{q_1^\sigma \to 0} \sum_{n \in \mathbb{Z}+\nu} \C[\mathcal{A}_4]{\theta_0,\theta_t,\sigma,\theta_1,\theta_\infty}{t}{j_1,j_2,n,j_3,j_4}{d},
    \end{equation}
    where $\mathcal{A}_4$ is given by (\ref{A_4}).
\end{proposition}
To see this, note that the coefficients $\mathcal{A}_4$ appear in blowup relations that use the invariant form of the partition function $\mathcal{F}_4$ (\ref{F-4-full}).
An advantage of this form is that the 1-loop part goes to $1$ in the limit $q_1^{\sigma} \to 0$. Moreover, as can be seen from the asymptotics of $\mathcal{N}$ factors in $\mathcal{F}_4^{\text{inst}}$,
\begin{equation}
    \lim\limits_{q_1^\sigma \to 0} \Z[\mathcal{F}_4^{\text{inst}}]{q_1^{\theta_0},q_1^{\theta_t},q_1^{\sigma},q_1^{\theta_1},q_1^{\theta_\infty}}{t,1} = \mathcal{F}_{4}^{\text{extra}}(q_1^{\theta_t},q_1^{\theta_1};q_1,q_2;t,1)^{-1}.
\end{equation}
These two observations ensure that
\begin{equation}
    \lim\limits_{q_1^\sigma \to 0} \Z[\mathcal{F}_4]{q_1^{\theta_0},q_1^{\theta_t},q_1^{\sigma},q_1^{\theta_1},q_1^{\theta_\infty}}{t,1} = 1.
\end{equation}
Note that $f(\pmb{\theta},t|\pmb{j},d)$ does not depend on $\sigma$.
Taking the $q_1^\sigma \to 0$ limit of the both sides of a blowup relation written in invariant form we obtain (\ref{f-a1-ans}).
Proposition \ref{proposition:f-a1-ans} opens the way to classifying all possible blowup relations.
\begin{proposition} \label{proposition:all-A4}
    There are no blowup relations of the form (\ref{blowup4}) beyond those presented in Table \ref{A1-table}.
\end{proposition}
To see this, observe the $\sigma$-dependence of $\mathcal{A}_4$ (\ref{A_4}), which is a simple exponential
\begin{equation}
    \C[\mathcal{A}_4]{\theta_0,\theta_t,\sigma,\theta_1,\theta_\infty}{t}{j_1,j_2,n,j_3,j_4}{d} \sim q_1^{\sigma\big(\big(n+\frac{j_2+j_3+2d}{4}\big)^2 + \frac{1}{4} - \frac{1}{4} Q^{(D_5)}(\pmb{j},d)\big)},
\end{equation}
where $Q^{(D_5)}(\pmb{j},d)$ is defined by (\ref{a1-quadratic-form}).
Since the lhs of a blowup relation does not depend on $\sigma$, there are two options:

1. Some $\mathcal{A}_4(n)$ diverge in the limit $q_1^\sigma \to 0$, but these divergencies cancel in the sum (\ref{f-a1-ans}), providing a finite limit.

2. There are no divergencies, all powers of $q_1^{\sigma}$ are positive, and the following inequality holds for all $n$:
\begin{equation} \label{a1-inequality}
    \Big(n+\frac{j_2+j_3+2d}{4}\Big)^2 + \frac{1}{4} - \frac{1}{4} Q^{(D_5)}(\pmb{j},d) \geq 0,
\end{equation}

We consider the option one first. Let $n^\ast$ be the smallest value of $n$ for which the lhs of (\ref{a1-inequality}) is negative and minimal.
Then this power of $q_1^{\sigma}$ should cancel the next one, corresponding to $n^\ast+1$. Since the dependence of $n$ in (\ref{a1-inequality})
is quadratic, the condition that these two exponentials coincide gives a linear equation on $n^\ast$
\begin{equation}
    n^\ast = -\frac{1+d}{2}-\frac{j_2+j_3}{4}.
\end{equation}
Once we have the value of $n^\ast$, we substitute it to the equation $\mathcal{A}_4(n^\ast) + \mathcal{A}_4(n^\ast + 1) = 0$. From the formula (\ref{A_4})
we obtain that the only solution is $(\pmb{j},d)=0$. For these values of $(\pmb{j},d)$ (\ref{a1-inequality}) holds. Therefore, option one never occurs.

Now assume that (\ref{a1-inequality}) holds for all $n$. It follows that
\begin{equation}
    Q^{(D_5)}(\pmb{j},d) \leq 1+4 \text{Inf}_n \Big(n+\frac{j_2+j_3+2d}{4}\Big)^2.
\end{equation}
Since $n$ increments by one, any number lies within a distance of at most $1/2$ from some value of $n$, which provides
\begin{equation} \label{a1-weight-criterium}
    Q^{(D_5)}(\pmb{j},d) \leq 2.
\end{equation}
A finite number of points of the lattice $\mathbb{Z}^5$ satisfy this condition.
Those that also satisfy the parity condition (\ref{a1-parity}) are precisely the tuples given in Table \ref{A1-table}.
We remark that two-term relations (\ref{blowup4-2-term}) allow weights that do not satisfy (\ref{a1-weight-criterium}) because of
cancellation between the two terms. In particular, for the given example of such relation (\ref{blowup4-2-term}), 
the quadratic form takes on negative values exactly once for each tuple of $(\pmb{j},d) = (-1, -1, 1, \pm 1, -1)$, and the corresponding terms cancel in the weighted sum.

\subsection{\texorpdfstring{$A_2$ relations}{A2}}
We now proceed to the $A_2$ quiver blowup relations.
\begin{conjecture} \label{conjecture:A2}
We have found $580$ blowup relations for a (general) $A_2$ quiver partition function
    \begin{equation} \label{blowup5}
        \begin{gathered}
            f(\pmb{\theta},t,y|\pmb{j},d,r) \times \Z[\mathcal{F}_5^\text{inst}]{q_1^{\theta_0}, q_1^{\theta_t}, q_1^{\sigma_1}, q_1^{\theta_1}, q_1^{\sigma_2}, q_1^{\theta_y}, q_1^{\theta_\infty}}{q_1, q_2; t, 1, y} = \\
            = \sum_{(n_1,n_2)\in\mathbb{Z}^2+(\nu_1,\nu_2)} \C[\mathcal{C}_5]{\theta_0, \theta_t, \sigma_1, \theta_1, \sigma_2, \theta_y, \theta_\infty}{t, y^{-1}}{j_1, j_2, n_1, j_3, n_2, j_4, j_5}{d,-r} \times \\
            \times \Z[\mathcal{F}_5^\text{inst}]{q_1^{\theta_0+j_1}, q_1^{\theta_t+j_2}, q_1^{\sigma+2n_1}, q_1^{\theta_1+j_3}, q_1^{\sigma_2+2n_2}, q_1^{\theta_y+j_4}, q_1^{\theta_\infty+j_5}}{q_1, q_1^{-1}q_2; t q_1^d, 1, yq_1^r} \times \\
            \times \Z[\mathcal{F}_5^\text{inst}]{q_1^{\theta_0}q_2^{j_1}, q_1^{\theta_t}q_2^{j_2}, q_1^{\sigma_1}q_2^{2n_1}, q_1^{\theta_1}q_2^{j_3}, q_1^{\sigma_2}q_2^{2n_2},q_1^{\theta_y}q_2^{j_4},q_1^{\theta_\infty}q_2^{j_5}}{q_1q_2^{-1}, q_2; t q_2^d, 1, yq_2^r},
        \end{gathered}
    \end{equation}
where every $(\pmb{j},d,r)$ belongs to $\mathbb{Z}^7$, and
\begin{align}
    \nu_1 &= \frac{1}{2}\Big(j_1+j_2+1 \mod 2\Big) \\
    \nu_2 &= \frac{1}{2}\Big(j_4+j_5+1 \mod 2\Big) \\
\end{align}
For the definition of $\mathcal{C}_5$ see (\ref{C_5}).
The following table contains some examples of these relations \footnote{It contains the highest weight representative of each orbit; for the definition of orbits and weights see below.}
\begin{table}[H]
    \centering
    \begin{tblr}{width=1\linewidth, cells={valign=m,halign=c}, row{1,2,3,4,5,6,7,8,9,10}={rowsep=8pt}, colspec={X[-1]X[-1]X[-1]}, vlines}
        \hline
        $(\nu_1,\nu_2)$ & $f(\pmb{\theta},t,y|\pmb{j},d,r)$ & $(\pmb{j},d,r)$ \\
        \hline
        $(0,\frac{1}{2})$ & $0$ & ${(1, 0, 0, 0, 0, 0, 0)}$ \\
        \hline
        $(\frac{1}{2},0)$ & $\mathcal{C}_{1/2,0}+\mathcal{C}_{-1/2,0}$ & $(2, 0, 0, 0, 1, 0, 0)$ \\
        \hline
        $(\frac{1}{2},0)$ & $0$ & $(0, 0, 0, 1, 0, 0, 0)$ \\
        \hline
        $(\frac{1}{2},\frac{1}{2})$ & $\frac{(\mathcal{C}_{1/2,1/2}+\mathcal{C}_{1/2,-1/2}+\mathcal{C}_{-1/2,1/2}+\mathcal{C}_{-1/2,-1/2})}{(1-y^{-1}q_1^{\theta_1})(1-ty^{-1}q_1^{\theta_t+\theta_1})}$ & $(0, 0, 1, 0, 0, 0, 0)$ \\
        \hline
        $(\frac{1}{2},0)$ & $\mathcal{C}_{1/2,0}+\mathcal{C}_{-1/2,0}$ & $(2, 0, 0, 1, 0, 0, 0)$ \\
        \hline
        $(0,\frac{1}{2})$ & $\mathcal{C}_{0,1/2}+\mathcal{C}_{0,-1/2}$ & $(1, 0, 0, 1, 1, 0, 0)$ \\
        \hline
        $(0,0)$ & $\frac{1-ty^{-1}\gamma^{-3}q_1^{(\theta_0+\theta_t+3\theta_1+\theta_y+\theta_\infty)/2}}{(1-y^{-1}q_1^{\theta_1})(1-t y^{-1}q_1^{\theta_t+\theta_1})}$ & $(1, 0, 1, 0, 1, 0, 0)$ \\
        \hline
        $(\frac{1}{2},0)$ & $\frac{1-ty^{-1}\gamma^{-2}q_1^{\theta_t+\theta_1+\theta_y}}{(1-tq_1^{\theta_t})(1-ty^{-1}q_1^{\theta_t+\theta_1})}(\mathcal{C}_{1/2,0}+\mathcal{C}_{-1/2,0})$ & $(1, 1, 0, 1, 0, 0, 0)$ \\
        \hline
        $(0, \frac{1}{2})$ & $\frac{ty^{-1/4}\gamma^{-2}q_1^{(2\theta_0+2\theta_t+3\theta_1+\theta_y)/4}}{(1-tq_1^{\theta_t})(1-ty^{-1}q_1^{\theta_t+\theta_1})}$ & $(1, 0, 0, 0, 0, 1, 0)$ \\
        \hline
        $(0,\frac{1}{2})$  & {$\frac{(1-ty^{-1}\gamma^{-2}q_1^{\theta_t+\theta_1})(1-ty^{-1}\gamma^{-2}q_1^{\theta_t+\theta_1+\theta_y})(\mathcal{C}_{0,1/2}+\mathcal{C}_{0,-1/2})}{(1-y^{-1}q_1^{\theta_1})(1-tq_1^{\theta_t})(1-ty^{-1}q_1^{\theta_t+\theta_1})(1-ty^{-1}q_1^{\theta_t+\theta_1+1})(1-ty^{-1}q_1^{\theta_t+\theta_1}q_2)}$} & $(0, 1, 0, 0, 0, 0, -1)$ \\
        \hline
    \end{tblr}
    \caption{$A_2$ relations \label{A2-relations}}
\end{table}
\end{conjecture}
\begin{proposition} \label{proposition:A2-relations-symmetries}
There is a $\widetilde{W}(A_1 \times A_1 \times A_5)$ action on the set of blowup relations. Compared to $W(A_1 \times A_1 \times A_5)$ it is extended by the two diagram automorphisms: $\sigma_1$ ($A_5$ reflection) and $\sigma_2$ (swaps the two copies of $A_1$). It is given by
\begin{equation} \label{A2:t1}
    t_1: (j_1,j_2,j_3,j_4,j_5,d,r) \mapsto (-j_1,j_2,j_3,j_4,j_5,d,r),
\end{equation}
\begin{equation} \label{A2:t2}
    t_2: (j_1,j_2,j_3,j_4,j_5,d,r) \mapsto (j_1,j_2,j_3,j_4,-j_5,d,r),
\end{equation}
\begin{equation} \label{A2:s1}
    s_1: (j_1,j_2,j_3,j_4,j_5,d,r) \mapsto (j_1,-j_2,j_3,j_4,j_5,d+j_2,r),
\end{equation}
\begin{equation} \label{A2:s2}
    s_2: (j_1,j_2,j_3,j_4,j_5,d,r) \mapsto (j_1,j_2+d,j_3+d,j_4,j_5,-d,r),
\end{equation}
\begin{equation} \label{A2:s3}
    s_3: (j_1,j_2,j_3,j_4,j_5,d,r) \mapsto (j_1,j_2,-j_3,j_4,j_5,d+j_3,r-j_3),
\end{equation}
\begin{equation} \label{A2:s4}
    s_4: (j_1,j_2,j_3,j_4,j_5,d,r) \mapsto (j_1,j_2,j_3-r,j_4-r,j_5,d,-r),
\end{equation}
\begin{equation} \label{A2:s5}
    s_5: (j_1,j_2,j_3,j_4,j_5,d,r) \mapsto (j_1,j_2,j_3,-j_4,j_5,d,r-j_4),
\end{equation}
\begin{equation} \label{A2:sigma1}
    \sigma_1: (j_1,j_2,j_3,j_4,j_5,d,r) \mapsto (j_1,j_4,j_3,j_2,j_5,-r,-d),
\end{equation}
\begin{equation} \label{A2:sigma2}
    \sigma_2: (j_1,j_2,j_3,j_4,j_5,d,r) \mapsto (j_5,j_2,j_3,j_4,j_1,d,r).
\end{equation}
$f$-factors in the lhs of (\ref{blowup5}) transform according to
\begin{equation} \label{A2:f_action}
    \begin{gathered}
        f(\pmb{\theta},t,y|w(\pmb{j},d,r)) = W\Bigg(f(\pmb{\theta},t,y|\pmb{j},d,r) \C[\mathcal{A}_5]{\theta_0,\theta_t,\sigma_1,\theta_1,\sigma_2,\theta_y,\theta_\infty}{t,y^{-1}}{j_1,j_2,n_1,j_3,n_2,j_4,j_5}{d,-r}^{-1}\Bigg) \\
        \times w\Bigg(\C[\mathcal{A}_5]{\theta_0,\theta_t,\sigma_1,\theta_1,\sigma_2,\theta_y,\theta_\infty}{t,y^{-1}}{j_1,j_2,n_1,j_3,n_2,j_4,j_5}{d,-r}\Bigg),
    \end{gathered}
\end{equation}
where
    $\mathcal{A}_5$ is defined in (\ref{A_5}), $w$ is $t_i$ or $s_i$, and $W$ is the corresponding $T_i$ or $S_i$, acting by (\ref{A2:T1}--\ref{A2:S5}).

\end{proposition}

In our coordinates $(j_2,j_3,j_4,d,r)$ the Killing form is
\begin{equation} \label{A5-Killing-form}
    K^{(A_5)} = \begin{pmatrix}
        \frac{5}{6} & \frac{1}{2} & \frac{1}{6} & \frac{2}{3} & -\frac{1}{3} \\
        \frac{1}{2} & \frac{3}{2} & \frac{1}{2} & 1 & -1 \\
        \frac{1}{6} & \frac{1}{2} & \frac{5}{6} & \frac{1}{3} & -\frac{2}{3} \\
        \frac{2}{3} & 1 & \frac{1}{3} & \frac{4}{3} & -\frac{2}{3} \\
        -\frac{1}{3} & -1 & -\frac{2}{3} & -\frac{2}{3} & \frac{4}{3}
    \end{pmatrix}.
\end{equation}
The corresponding invariant quadratic forms are
\begin{equation}
    \begin{gathered}
        Q^{(A_5)}(j_2,j_3,j_4,d,r) = \frac{j_2^2}{6} + \frac{j_3^2}{2} + \frac{j_4^2}{6} + \frac{2(d+r)^2}{9} + \frac{(j_2+j_4)^2}{6} + \\
        + \frac{1}{2}\Big(j_2+j_3+\frac{2}{3} \Big(2d-r\Big)\Big)^2 + \frac{1}{2} \Big(j_3+j_4+\frac{2}{3}\Big(d-2r\Big) \Big)^2,
    \end{gathered}
\end{equation}
and
\begin{equation} \label{a2-quadratic-form}
    Q^{(A_1 \times A_1 \times A_5)}(j_1,j_2,j_3,j_4,j_5,d,r) = Q^{(A_5)}(j_2,j_3,j_4,d,r) + \frac{j_1^2}{2} + \frac{j_5^2}{2}.
\end{equation}
The fundamental weights and simple roots of $\mathfrak{sl}_6$ are
\begin{equation}
    \begin{aligned}
        \omega_1 &= (1, 0, 0, 0, 0), \\
        \omega_2 &= (0, 0, 0, 1, 0), \\
        \omega_3 &= (0, 1, 0, 0, 0), \\
        \omega_4 &= (0, 0, 0, 0, -1), \\
        \omega_5 &= (0, 0, 1, 0, 0),
    \end{aligned}
    \quad\quad
    \begin{aligned}
        \alpha_1 &= (2, 0, 0, -1, 0), \\
        \alpha_2 &= (-1, -1, 0, 2, 0), \\
        \alpha_3 &= (0, 2, 0, -1, 1), \\
        \alpha_4 &= (0, -1, -1, 0, -2), \\
        \alpha_5 &= (0, 0, 2, 0, 1),
    \end{aligned}
\end{equation}
and the fundamental weights of $\mathfrak{sl}_2 \oplus \mathfrak{sl}_2$ in the coordinates $(j_1,j_5)$ are $(1,0)$ and $(0,1)$.

The full set of 580 relations splits into the orbits of the extended Weyl group action. Their cardinalities are
\begin{equation}
    580 = 4+8+12+20+2\times48+80+3\times120.
\end{equation}
If one restricts to the action of $W(A_5)$ these cardinalities become
\begin{equation} \label{W(A5)-cardinalities}
    580 = 12 \times 1 + 18 \times 6 + 8 \times 15 + 5 \times 20  + 4 \times 30  + 2 \times 60.
\end{equation}
Let $\Omega$ be the set of the tuples $(\pmb{j},d,r)$ that appear in the blowup relations.
To describe the orbits we introduce the following character of $\mathfrak{sl}_2 \oplus \mathfrak{sl}_2 \oplus \mathfrak{sl}_6$
\begin{equation}
    \chi = \sum_{w\in\Omega} (y_1,x_1,x_2,x_3,y_2,x_4,x_5)^w.
\end{equation}
We aim to expand it into a sum of products of the characters $\chi^{11}(\pmb{y})$ of $\mathfrak{sl}_2 \oplus \mathfrak{sl}_2$ and $\chi^5(\pmb{x})$ of $\mathfrak{sl}_6$.
The full character expands as
\begin{equation} \label{A2-character}
    \begin{gathered}
        \chi = \chi^{11}_{(1,2)} + \chi^{11}_{(2,1)} + (\chi^{11}_{(0,0)}+\chi^{11}_{(1,1)}+\chi^{11}_{(2,0)}+\chi^{11}_{(0,2)}-2)(\chi^5_{\omega_1}+\chi^5_{\omega_5}) + \\
        + (\chi^{11}_{(1,0)} + \chi^{11}_{(0,1)})(\chi^5_{\omega_2}+\chi^5_{\omega_4}) + (\chi^{11}_{(0,0)}+\chi^{11}_{(1,1)})\chi^5_{\omega_3} + \\
        + (\chi^{11}_{(1,0)}+\chi^{11}_{(0,1)})(\chi^5_{\omega_1+\omega_5}-5) + (\chi^5_{\omega_1+\omega_4}-4\chi^5_{\omega_5}) + (\chi^5_{\omega_2+\omega_5}-4\chi^5_{\omega_1}).
    \end{gathered}
\end{equation}
This expansion refines the formula (\ref{W(A5)-cardinalities}).

Similarly to the $A_1$ case, there is a formula for $f$ factors in terms of $\mathcal{A}_5$ coefficients.
\begin{proposition} \label{proposition:f-a2-ans}
    Assume that Conjecture \ref{conjecture:A2} holds.
    Then the factors $f(\pmb{\theta},t,y|\pmb{j},d,r)$ in (\ref{blowup5}) are given by
    \begin{equation} \label{f-a2-ans}
        \begin{gathered}
            f(\pmb{\theta},t,y|\pmb{j},d,r) = \\
            = \lim\limits_{q_1^\sigma \to 0} \sum_{(n_1,n_2) \in \mathbb{Z}^2+(\nu_1,\nu_2)} \C[\mathcal{A}_5]{\theta_0, \theta_t, \sigma, \theta_1, \sigma^\ast+\sigma, \theta_y, \theta_\infty}{t, y^{-1}}{j_1, j_2, n_1, j_3, n_2, j_4, j_5}{d,-r},
        \end{gathered}
    \end{equation}
    where $\mathcal{A}_5$ is given by (\ref{A_5}).
\end{proposition}
As in the $A_1$ case, the statement follows from the normalization of the invariant partition function that gives
\begin{equation}
    \lim\limits_{q_1^\sigma \to 0} \Z[\mathcal{F}_5]{q_1^{\theta_0}, q_1^{\theta_t}, q_1^{\sigma}, q_1^{\theta_1}, q_1^{\sigma^\ast+\sigma}, q_1^{\theta_y}, q_1^{\theta_\infty}}{q_1, q_2; t, 1, y} = 1.
\end{equation}
Now we use (\ref{f-a2-ans}) to classify $A_2$ blowup relations.
\begin{proposition}
    There are no blowup relations of the form (\ref{blowup5}) beyond the $580$ relations generated by the Weyl group action (\ref{A2:t1}-\ref{A2:f_action}) from the relations presented in Table \ref{A2-relations}.
\end{proposition}
This result is analogous to Proposition \ref{proposition:all-A4}. We use (\ref{f-a2-ans}) and examine the $\sigma$-dependence
\begin{equation}
    \C[\mathcal{A}_5]{\theta_0, \theta_t, \sigma, \theta_1, \sigma^\ast+\sigma, \theta_y, \theta_\infty}{t, y^{-1}}{j_1, j_2, n_1, j_3, n_2, j_4, j_5}{d,-r} \sim q_1^{\sigma Q(n_1,n_2)},
\end{equation}
\begin{equation}
    \begin{gathered}
        Q(n_1,n_2) = \Big(n_1-\frac{n_2}{2}+\frac{2d+j_2+j_3}{4}\Big)^2 + \frac{3}{4}\Big(n_2+\frac{j_2+3j_3+2j_4+2d-4r}{6}\Big)^2 + \\
        + \frac{3}{8} - \frac{1}{4} Q^{(A_1\times A_1\times A_5)}(\pmb{j},d,r),
    \end{gathered}
\end{equation}
where $Q^{(A_1\times A_1\times A_5)}$ is defined by (\ref{a2-quadratic-form}).
As in the $A_1$ case, a priori, two options are possible: either singularities are present but cancel between different $n_1, n_2$, or no singularities are present.
To study the possible cancellation of singularities, note that
\begin{equation}
    Q(n_1,n_2) = q(n_1-c_1,n_2-c_2) + c_3,
\end{equation}
where $c = (c_1, c_2)$ and $c_3$ are $n$-independent, and $q$ is the quadratic part
\begin{equation}
    q(n_1,n_2) = n_1^{2} - n_1 n_2 + n_2^{2}.
\end{equation}
Denote the associated scalar product by $\braket{}$, 
\begin{equation}
    \braket{u,v} = \frac{1}{2}\Big(q(u+v)-q(u)-q(v)\Big).
\end{equation}
Let $n=(n_1,n_2)$ and $n' = (n_1',n_2')$ be two distinct points where $Q$ attains its minimum value on the lattice.
Then for any $v \in \mathbb{Z}^2$
\begin{equation}
    q(n-c-v) \geq q(n-c) \implies \braket{n-c,v} \leq \frac{1}{2}q(v)
\end{equation}
\begin{equation}
    q(n'-c+v) \geq q(n'-c) \implies -\braket{n'-c,v} \leq \frac{1}{2}q(v)
\end{equation}
Combining the last two,
\begin{equation}
    \braket{n-n',v} \leq q(v),
\end{equation}
and therefore,
\begin{equation}
    q(n-n'-2v) = q(n-n') + 4q(v)-4\braket{n-n',v} \geq q(n-n')
\end{equation}
Therefore, $n-n'$ is a minimum of the form $q$ on the coset $[n-n'] = n-n' +2\mathbb{Z}^2 \subset \mathbb{Z}^2$. 
There are four possible cosets, $[(0,0)], [(0,1)], [(1,0)]$, and $[(1,1)]$. The minimal values of $q$ in these classes 
correspond to the vectors $(0,0)$, $\pm(0,1)$, $\pm(1,0)$, and $\pm(1,1)$ respectively. 
Therefore, relabelling the points if needed, cancellations could happen in pairs $((n_1,n_2), (n_1+1,n_2))$, $((n_1,n_2), (n_1,n_2+1))$ and $((n_1,n_2), (n_1+1,n_2+1))$, 
or triples $((n_1,n_2), (n_1,n_2+1), (n_1+1,n_2+1))$ and $((n_1, n_2), (n_1+1, n_2), (n_1+1,n_2+1))$\footnote{Cancellation is not possible between $(n_1+1,n_2)$ and $(n_1,n_2+1)$ because $q(1,-1)=3 > 1 = q(1,1)$, therefore no pair or triple contains both.}.
Using the formula for $\mathcal{A}_5$, (\ref{A_5}), we check that no cancellation of any such pair or triple is possible. 
To see this, it is enough to compare the asymptotics of $\mathcal{A}_5$ at these adjacent points in every parameter 
$q_1^{\pmb{\theta}}, t, y, q_1^{\sigma}$, and $q_1^{\sigma^\ast}$.
From the requirement that these asymptotics coincide it is possible to solve for $\pmb{j}, d, r, n_1$ and $n_2$, 
they are fixed up to just one parameter.
Then this solution is substituted back into the sum of $\mathcal{A}_5$, and the asymptotics in $q_1^{\sigma_\ast}$ 
is checked again, which allows solving for that parameter.
The final configuration is a non-singular one, that is $Q(n_1,n_2,\pmb{j}, d, r) > 0$. Therefore, as in the $A_1$ 
case, the cancellation of singularities never occurs.
Therefore, the quadratic form $Q$ should be nonnegative, and one has 
\begin{equation}
    \begin{gathered}
        \text{Inf}_{n_1,n_2} \Big(n_1-\frac{n_2}{2}+\frac{2d+j_2+j_3}{4}\Big)^2 + \frac{3}{4}\Big(n_2+\frac{j_2+3j_3+2j_4+2d-4r}{6}\Big)^2 + \\
        + \frac{3}{8} - \frac{1}{4} Q^{(A_1\times A_1\times A_5)}(\pmb{j},d,r) \geq 0,
    \end{gathered}
\end{equation}
Since $n_1$ and $n_2$ increment by one, the infimum cannot be larger than $1/4+3/16=7/16$, and we obtain for the possible $(\pmb{j},d,r)$
\begin{equation}
    Q^{(A_1\times A_1\times A_5)}(\pmb{j},d,r) \leq \frac{13}{4}.
\end{equation}
All such $(\pmb{j},d,r) \in \mathbb{Z}^7$, satisfying the parity condition $j_1+j_2+j_3+j_4+j_5 \mod 2 = 1$, $580$ in total, are obtained by the Weyl group action from those presented in Table \ref{A2-relations}.
\subsection{\texorpdfstring{Two-term $A_2$ relations}{Two-term-A2}}
There is another type of blowup relations for the $A_2$ quiver partition functions, which we call two-term $A_2$ relations. Here we report some of these relations, which we will use in the derivation of the $q$-Painlev\'e VI bilinear relations. Similarly to (\ref{blowup4-2-term}), they include a sum of two terms in the rhs. We indicate it by a sum over $\omega$, a two-element set of tuples $(\pmb{j},d,r)$.
\begin{conjecture} \label{conjecture:two-term-A2}
Two-term $A_2$ blowup relations
\begin{equation} \label{Two-term-A2}
        \begin{gathered}
            \Z[\mathcal{F}_5^\text{inst}]{q_1^{\theta_0}, q_1^{\theta_t}, q_1^{\sigma_1}, q_1^{\theta_1}, q_1^{\sigma_2}, q_1^{\theta_y}, q_1^{\theta_\infty}}{q_1, q_2; t, 1, y} = \\
            = \sum_{(\pmb{j},d,r)\in\omega} g_{\omega}(\pmb{\theta},t,y|\pmb{j},d,r) \times \\
            \times \sum_{(n_1,n_2)\in\mathbb{Z}^2+(\nu_1^{(\pmb{j})},\nu_2^{(\pmb{j})})} \C[\mathcal{C}_5]{\theta_0, \theta_t, \sigma_1, \theta_1, \sigma_2, \theta_y, \theta_\infty}{t, y^{-1}}{j_1, j_2, n_1, j_3, n_2, j_4, j_5}{d,-r} \times \\
            \times \Z[\mathcal{F}_5^\text{inst}]{q_1^{\theta_0+j_1}, q_1^{\theta_t+j_2}, q_1^{\sigma+2n_1}, q_1^{\theta_1+j_3}, q_1^{\sigma_2+2n_2}, q_1^{\theta_y+j_4}, q_1^{\theta_\infty+j_5}}{q_1, q_1^{-1}q_2; t q_1^d, 1, yq_1^r} \times \\
            \times \Z[\mathcal{F}_5^\text{inst}]{q_1^{\theta_0}q_2^{j_1}, q_1^{\theta_t}q_2^{j_2}, q_1^{\sigma_1}q_2^{2n_1}, q_1^{\theta_1}q_2^{j_3}, q_1^{\sigma_2}q_2^{2n_2},q_1^{\theta_y}q_2^{j_4},q_1^{\theta_\infty}q_2^{j_5}}{q_1q_2^{-1}, q_2; t q_2^d, 1, yq_2^r}.
        \end{gathered}
\end{equation}
Now $\nu_1^{(\pmb{j})}$ and $\nu_2^{(\pmb{j})}$ carry an index $\pmb{j}$ because they may be different for the two tuples of $\pmb{j}$ in a given $\omega$. As before,
\begin{equation}
    \begin{gathered}
        \nu_1^{(\pmb{j})} = \frac{1}{2}\Big(j_1+j_2+1 \mod 2\Big), \\
        \nu_2^{(\pmb{j})} = \frac{1}{2}\Big(j_4+j_5+1 \mod 2\Big).
    \end{gathered}
\end{equation}
\end{conjecture}
The following table contains $(\omega, g_\omega)$ for $4$ relations of this form.
\begin{table}[H]
    \centering
    \begin{tblr}{width=1\linewidth, cells={valign=m,halign=c}, row{1,2,3,4,5,6,7,8,9,10}={rowsep=5pt}, colspec={X[-1]X[-1]}, vlines}
        \hline
        $g_\omega$ & $\omega$ \\
        \hline
        $\Big\{\frac{y^{1/4}q_1^{(3-\theta_1+3\theta_y)/4}q_2^{3/4}(1-tq_1^{\theta_1})}{(1-ty^{-1}q_1^{\theta_1-1}q_2^{-1})(1-q_1^{\theta_y+1}q_2)},$ & $\{(1, 0, 0, 2, 0, -1, 1),$ \\
        $\frac{t^{3/4}y^{1/4}q_1^{(1-2\theta_0+\theta_t+3\theta_y)/4}q_2^{1/4}(1-tq_1^{\theta_1})(1-y^{-1}q_1^{\theta_1-1}q_2^{-1})}{(1-ty^{-1}q_1^{\theta_1-1}q_2^{-1})(1-q_1^{\theta_y+1}q_2)}\Big\}$ & $(0, 0, -1, 2, 0, 0, 1)\}$ \\
        \hline
        $\Big\{\frac{y^{1/4}q_1^{(3+4\theta_0-\theta_1+3\theta_y)/4}q_2^{3/4}(1-tq_1^{\theta_1})}{(1-ty^{-1}q_1^{\theta_1-1}q_2^{-1})(q_1^{\theta_0}-1)(1-q_1^{\theta_y+1}q_2)},$ & $\{(1, 0, 0, 2, 0, -1, 1),$ \\
        $-\frac{y^{1/4}q_1^{(3-\theta_1+3\theta_y)/4}q_2^{3/4}(1-tq_1^{\theta_1})}{(1-ty^{-1}q_1^{\theta_1-1}q_2^{-1})(q_1^{\theta_0}-1)(1-q_1^{\theta_y+1}q_2)}\Big\}$ & $(-1, 0, 0, 2, 0, -1, 1)\}$ \\
        \hline
        $\Big\{\frac{y^{1/4}q_1^{(3-\theta_1+3\theta_y)/4}q_2^{3/4}(1-tq_1^{\theta_1})}{(1-tq_1^{(-3-\theta_0+\theta_t+\theta_1-\theta_y+\theta_\infty)/2}q_2^{-3/2})(1-ty^{-1}q_1^{\theta_1-1}q_2^{-1})(1-q_1^{\theta_y+1}q_2)},$ & $\{(1, 0, 0, 2, 0, -1, 1),$ \\
        $-\frac{t^{3/4}q_1^{(-1-2\theta_0+\theta_t+\theta_1+2\theta_y-2\theta_\infty)/4}q_2^{-1/4}(1-tq_1^{\theta_1})}{(1-tq_1^{(-3-\theta_0+\theta_t+\theta_1-\theta_y+\theta_\infty)/2}q_2^{-3/2})(1-ty^{-1}q_1^{\theta_1-1}q_2^{-1})(1-q_1^{\theta_y+1}q_2)}\Big\}$ & $(0, 0, 0, 2, -1, -1, 1)\}$ \\
        \hline
        $\Big\{\frac{y^{1/4}q_1^{(3-\theta_1+3\theta_y)/4}q_2^{3/4}(1-tq_1^{\theta_1})}{(1-tq_1^{(-3-\theta_0+\theta_t+\theta_1-\theta_y-\theta_\infty)/2}q_2^{-3/2})(1-ty^{-1}q_1^{\theta_1-1}q_2^{-1})(1-q_1^{\theta_y+1}q_2)},$ & $\{(1, 0, 0, 2, 0, -1, 1),$ \\
        $-\frac{t^{3/4}q_1^{(-1-2\theta_0+\theta_t+\theta_1+2\theta_y+2\theta_\infty)/4}q_2^{-1/4}(1-tq_1^{\theta_1})}{(1-tq_1^{(-3-\theta_0+\theta_t+\theta_1-\theta_y-\theta_\infty)/2}q_2^{-3/2})(1-ty^{-1}q_1^{\theta_1-1}q_2^{-1})(1-q_1^{1+\theta_y}q_2)}\Big\}$ & $(0, 0, 0, 2, 1, -1, 1)\}$ \\
        \hline
    \end{tblr}
    \caption{Examples of two-term $A_2$ relations \label{A2-two-term}}
\end{table}
More relations can be generated by the Weyl group action. We do not study them and focus on these four as we will use them in Section \ref{tau-functions-bilinear-relations} to derive bilinear relations for the q-Painlev\'e VI tau functions.
\subsection{\texorpdfstring{Higgsed $A_2$ relations}{Higgsed-A2}}
In order to derive the blowup relations relevant to $q$-Painlev\'e VI we now consider the Higgsed $A_2$ relations. 
In terms of the gauge theory Higgsing is a way to introduce half-BPS surface defect \cite{N_2020} by tuning some of the 
fundamental masses and a Coulomb modulus. 
By AGT these partition functions correspond to a $q$-deformation of the conformal blocks with one of the primary fields 
being degenerate. In the case we are considering here the corresponding degenerate field (recovered in the $4$d limit) 
is either the identity or the $(2,1)$ field. We deduce the corresponding special blowup relations from the general $A_2$ 
relations of the previous section. To do this we set $q_1^{\theta_t}$, $q_1^{\theta_1}$ or $q_1^{\theta_y}$ to $(q_1^2q_2)^{\pm 1}$ 
(note that we rename these variables in such a way that $y$ is always the position of the degenerate field) and set the 
adjacent $\theta$s to the values consistent with the OPE in the $4$d limit. We focus on the relations that have the 
corresponding $j=\pm2$ since these values ensure that one of the $\mathcal{F}_5$ partition functions on the rhs of 
a blowup relation has the same degenerate insertion as the one on the lhs while another one has the identity field 
at that position. The latter reduces to $\mathcal{F}_4$ by the following observation (which we checked in the first 
few orders of the instanton expansion)
\begin{equation}
    \Z[\mathcal{F}_5^\text{inst}]{q_1^{\theta_0}, q_1^{\theta_t}, q_1^{\sigma}, q_1^{\theta_1}, q_1^{\theta_\infty}, q_1^{-1}q_2^{-1}, q_1^{\theta_\infty}}{q_1, q_2; t, 1, y} = \Z[\mathcal{F}_4^\text{inst}]{q_1^{\theta_0}, q_1^{\theta_t}, q_1^{\sigma}, q_1^{\theta_1}, q_1^{\theta_\infty}}{q_1, q_2; t, 1},
\end{equation}
\begin{equation}
    \Z[\mathcal{F}_5^\text{inst}]{q_1^{\theta_0}, q_1 q_2, q_1^{\theta_0}, q_1^{\theta_t}, q_1^{\sigma}, q_1^{\theta_1}, q_1^{\theta_\infty}}{q_1, q_2; y, t, 1} = \Z[\mathcal{F}_4^\text{inst}]{q_1^{\theta_0}, q_1^{\theta_t}, q_1^{\sigma}, q_1^{\theta_1}, q_1^{\theta_\infty}}{q_1, q_2; t, 1},
\end{equation}
\begin{equation}
    \begin{gathered}
        \Z[\mathcal{F}_5^\text{inst}]{q_1^{\theta_0}, q_1^{\theta_t}, q_1^{\sigma}, q_1 q_2, q_1^{\sigma}, q_1^{\theta_1}, q_1^{\theta_\infty}}{q_1, q_2; t, y, 1} = \\
        = \frac{(q_1^{\theta_t} ty^{-1};q_1,q_2)_\infty (\gamma^{-4} ty^{-1};q_1,q_2)_\infty}{(\gamma^{-2} ty^{-1};q_1,q_2)_\infty (\gamma^{-2} q_1^{\theta_t} ty^{-1};q_1,q_2)_\infty} \Z[\mathcal{F}_4^\text{inst}]{q_1^{\theta_0}, q_1^{\theta_t}, q_1^{\sigma}, q_1^{\theta_1}, q_1^{\theta_\infty}}{q_1, q_2; q_1q_2 t, 1} = \\
        = \frac{(q_1^{\theta_t}ty^{-1};q_2)_\infty (q_1^{\theta_t+1}ty^{-1};q_1)_\infty}{(q_1q_2ty^{-1};q_2)_\infty (q_1^2q_2ty^{-1};q_1)_\infty} \Z[\mathcal{F}_4^\text{inst}]{q_1^{\theta_0}, q_1^{\theta_t}, q_1^{\sigma}, q_1^{\theta_1}, q_1^{\theta_\infty}}{q_1, q_2; q_1q_2 t, 1},
    \end{gathered}
\end{equation}
or,
\begin{equation}
    \begin{gathered}
        \Z[\mathcal{F}_5^\text{inst}]{q_1^{\theta_0}, q_1^{\theta_t}, q_1^{\sigma}, q_2, q_1^{\sigma}, q_1^{\theta_1}, q_1^{\theta_\infty}}{q_1, q_1^{-1}q_2; t, y, 1} = \\
        = \frac{(q_1^{\theta_t}ty^{-1};q_1^{-1}q_2)_\infty (q_1^{\theta_t+1}ty^{-1};q_1)_\infty}{(q_2ty^{-1};q_1^{-1}q_2)_\infty (q_1 q_2ty^{-1};q_1)_\infty} \Z[\mathcal{F}_4^\text{inst}]{q_1^{\theta_0}, q_1^{\theta_t}, q_1^{\sigma}, q_1^{\theta_1}, q_1^{\theta_\infty}}{q_1, q_1^{-1}q_2; q_2 t, 1},
    \end{gathered}
\end{equation}
A similar degeneration occurs for the $\mathcal{C}$ coefficients. It can be checked that $\mathcal{C}_5$ reduces to $\mathcal{C}_4$ up to some factor (\ref{C_5-C_4-reduction}--\ref{C_5-C_4-reduction-2}).
\begin{conjecture} \label{conjecture:A2-higgsed}
We have found the following blowup relations for the Higgsed $A_2$ quiver partition function. $h$ is a parameter taking values $\pm1$.
\begin{equation} \label{blowup5-higgsed-1}
    \begin{gathered}
        f(\pmb{\theta},t,y|\pmb{j},d,r) \times \Z[\mathcal{F}_5^\text{inst}]{q_1^{\theta_0}, q_1^{\theta_t}, q_1^{\sigma}, q_1^{\theta_1}, q_1^{\theta_\infty-1-h}, q_1^{-2}q_2^{-1}, q_1^{\theta_\infty-1}}{q_1, q_2; t, 1, y} = \\
        = \sum_{n\in\mathbb{Z}+\nu} \C[\mathcal{C}_4]{\theta_0,\theta_t, \sigma, \theta_1, \theta_\infty-1-h}{t}{j_1,j_2,n,j_3,j_4}{d} \times \\
        \times \Z[\mathcal{F}_4^\text{inst}]{q_1^{\theta_0+j_1}, q_1^{\theta_t+j_2}, q_1^{\sigma+2n}, q_1^{\theta_1+j_3}, q_1^{\theta_\infty-1-h+j_4}}{q_1, q_1^{-1}q_2; t q_1^d, 1} \times \\
        \times \Z[\mathcal{F}_5^\text{inst}]{q_1^{\theta_0}q_2^{j_1}, q_1^{\theta_t}q_2^{j_2}, q_1^{\sigma}q_2^{2n}, q_1^{\theta_1}q_2^{j_3}, q_1^{\theta_\infty-1-h}q_2^{j_4},q_1^{-2}q_2,q_1^{\theta_\infty-1}q_2^{j_4-h}}{q_1q_2^{-1}, q_2; t q_2^d, 1, yq_2^r},
    \end{gathered}
\end{equation}
where $f(\pmb{\theta},t,y|\pmb{j},d,r)$ are given in Table \ref{Higgsed_A_2-t1y},
\begin{equation} \label{blowup5-higgsed-2}
    \begin{gathered}
        f(\pmb{\theta},t,y|\pmb{j},d,r) \times \Z[\mathcal{F}_5^\text{inst}]{q_1^{\theta_0}, q_1^2q_2, q_1^{\theta_0-h}, q_1^{\theta_t}, q_1^{\sigma}, q_1^{\theta_1}, q_1^{\theta_\infty-1}}{q_1, q_2; y, t, 1} = \\
        = \sum_{n\in\mathbb{Z}+\nu} \C[\mathcal{C}_4]{\theta_0-h,\theta_t, \sigma, \theta_1,\theta_\infty-1}{t}{j_1,j_2, n,j_3,j_4}{d} \times \\
        \times \Z[\mathcal{F}_4^\text{inst}]{q_1^{\theta_0-h+j_1}, q_1^{\theta_t+j_2}, q_1^{\sigma+2n}, q_1^{\theta_1+j_3}, q_1^{\theta_\infty-1+j_4}}{q_1, q_1^{-1}q_2; t q_1^d, 1} \times \\
        \times \Z[\mathcal{F}_5^\text{inst}]{q_1^{\theta_0}q_2^{j_1-h}, q_1^2q_2^{-1}, q_1^{\theta_0-h}q_2^{j_1}, q_1^{\theta_t}q_2^{j_2}, q_1^{\sigma}q_2^{2n}, q_1^{\theta_1}q_2^{j_3}, q_1^{\theta_\infty-1}q_2^{j_4}}{q_1q_2^{-1}, q_2; yq_2^r, t q_2^d, 1},
    \end{gathered}
\end{equation}
where $f(\pmb{\theta},t,y|\pmb{j},d,r)$ are given in Table \ref{Higgsed_A_2-yt1},
\begin{equation} \label{blowup5-higgsed-3}
    \begin{gathered}
        f(\pmb{\theta},t,y|\pmb{j},d,r) \times \frac{(y q_1^r;q_1,q_1^{-1}q_2)_{\infty} (y q_1^{\theta_1+j_3+r};q_1,q_1^{-1}q_2)_{\infty}}{(y q_1^r q_2^{-1};q_1,q_1^{-1}q_2)_{\infty} (y q_1^{\theta_1+j_3+r}q_2;q_1,q_1^{-1}q_2)_{\infty}} \times \\
        \times \Z[\mathcal{F}_5^\text{inst}]{q_1^{\theta_0}, q_1^{\theta_t}, q_1^{\sigma}, q_1^{-2}q_2^{-1},q_1^{\sigma+h}, q_1^{\theta_1}, q_1^{\theta_\infty-1}}{q_1, q_2; t, y, 1} = \\
        = \sum_{n\in\mathbb{Z}+\nu} \C[\mathcal{C}_5]{\theta_0,\theta_t, \sigma, -2-\frac{\varepsilon_2}{\varepsilon_1}, \sigma+h, \theta_1, \theta_\infty-1}{t y^{-1},y}{j_1,j_2,n,2,n-\frac{h}{2},j_3,j_4}{d-r,r} \times \\
        \times \Z[\mathcal{F}_4^\text{inst}]{q_1^{\theta_0+j_1}, q_1^{\theta_t+j_2}, q_1^{\sigma+2n}, q_1^{\theta_1+j_3},q_1^{\theta_\infty-1+j_4}}{q_1, q_1^{-1}q_2; t q_1^d q_2^{-1},1} \times \\
        \times \Z[\mathcal{F}_5^\text{inst}]{q_1^{\theta_0}q_2^{j_1}, q_1^{\theta_t}q_2^{j_2}, q_1^{\sigma}q_2^{2n}, q_1^{-2}q_2, q_1^{\sigma+h}q_2^{2n-h},q_1^{\theta_1}q_2^{j_3},q_1^{\theta_\infty-1}q_2^{j_4}}{q_1q_2^{-1}, q_2; t q_2^d, y q_2^{r}, 1},
    \end{gathered}
\end{equation}
where $f(\pmb{\theta},t,y|\pmb{j},d,r)$ are given in Table \ref{Higgsed_A_2-ty1}.
\end{conjecture}
We conclude this subsection with a remark about the limits $t\to 0$ and $y \to \infty$  of the $A_2$ blowup relations.
The limit $t\to0$ of the Higgsed $A_2$ quiver partition function reduces to the $q$-hypergeometric function:
\begin{equation}
    \begin{gathered}
        \Z[\mathcal{F}_5^\text{inst}]{\ast, \quad \ast, \quad q_1^{\sigma}, q_1^{\theta_1}, q_1^{\theta_\infty-1-h}, q_1^{-2}q_2^{-1}, q_1^{\theta_\infty-1}}{q_1, q_2; 0, 1, y} = \\
        = \Z[\mathcal{F}_4^\text{inst}]{q_1^{\sigma}, q_1^{\theta_1}, q_1^{\theta_\infty-1-h}, q_1^{-2}q_2^{-1}, q_1^{\theta_\infty-1}}{q_1, q_2; 1, y} = \\
        = \tensor[_2]{\phi}{_1}\Bigg(\genfrac{}{}{0pt}{}{q_1^{(h(\theta_\infty-1)-\theta_1-\sigma)/2}q_2^{1/2}, \hspace{0.5em} q_1^{(h(\theta_\infty-1)-\theta_1+\sigma)/2}q_2^{1/2}}{q_1^{h(\theta_\infty-1)}q_2};q_2;q_1^{\theta_1-1}y^{-1}\Bigg).
    \end{gathered}
\end{equation}
In particular, there is a symmetry (lhs and rhs are equal to the same hypergeometric function)
\begin{equation} \label{hpg_symmetry}
    \begin{gathered}
        \Z[\mathcal{F}_5^\text{inst}]{\ast, \quad \ast, \quad q_1^{\sigma}, q_1^{\theta_1}, q_1^{\theta_\infty-1-h}, q_1^{-2}q_2^{-1}, q_1^{\theta_\infty-1}}{q_1, q_2; 0, 1, y} = \\
        = \Z[\mathcal{F}_5^\text{inst}]{\ast, \quad \ast, \quad q_1^{\sigma}, q_1^{\theta_1}, q_1^{\theta_\infty-1-h}q_2^h, q_1^{-2}q_2, q_1^{\theta_\infty-1}}{q_1 q_2^{-1}, q_2; 0, 1, q_2y}.
    \end{gathered}
\end{equation}
When $\nu=0$ and $t\to 0$  only $n=0$ contributes to the sum and (\ref{blowup5-higgsed-1}) with $\nu=0$ become (\ref{hpg_symmetry}).
The situation is different for half-integer relations. When $t\to0$ the sum on the right has two terms $(n=\pm1/2)$. In the limit $t\to0$ they reduce to Gauss's contiguous relations on $q$-hypergeometric functions.
We also note that in the limit $y\to\infty$ a general $A_2$ blowup relation (\ref{blowup5}) becomes an $A_1$ relation as in (\ref{blowup4}) or (\ref{blowup4-2-term}) depending on the parity of $2\nu_2$.

The two-term $A_2$ relations listed in Table \ref{A2-two-term} all have $j_4=2$. Therefore, they can be subject to Higgsing by setting $q_1^{\theta_y}=q_1^{-2}q_2^{-1}$, $\theta_\infty \to \theta_\infty-1$ and $q_1^{\sigma_2} = q_1^{\theta_\infty-1-h}$. Hence we get
\begin{equation} \label{Two-term-A2-Higgsed}
    \begin{gathered}
        \Z[\mathcal{F}_5^\text{inst}]{q_1^{\theta_0}, q_1^{\theta_t}, q_1^{\sigma}, q_1^{\theta_1}, q_1^{\theta_\infty-1-h}, q_1^{-2}q_2^{-1}, q_1^{\theta_\infty-1}}{q_1, q_2; t, 1, y} = \\
        = \sum_{(\pmb{j},d,r) \in \omega} g_{\omega}(\pmb{\theta},t,y|\pmb{j},d,r) \sum_{n\in\mathbb{Z}+\nu^{(\pmb{j})}} \C[\mathcal{C}_4]{\theta_0,\theta_t, \sigma, \theta_1, \theta_\infty-1-h}{t}{j_1,j_2,n,j_3,j_4}{d} \times \\
        \times \Z[\mathcal{F}_4^\text{inst}]{q_1^{\theta_0+j_1}, q_1^{\theta_t+j_2}, q_1^{\sigma+2n}, q_1^{\theta_1+j_3}, q_1^{\theta_\infty-1-h+j_4}}{q_1, q_1^{-1}q_2; t q_1^d, 1} \times \\
        \times \Z[\mathcal{F}_5^\text{inst}]{q_1^{\theta_0}q_2^{j_1}, q_1^{\theta_t}q_2^{j_2}, q_1^{\sigma}q_2^{2n}, q_1^{\theta_1}q_2^{j_3}, q_1^{\theta_\infty-1-h}q_2^{j_4},q_1^{-2}q_2,q_1^{\theta_\infty-1}q_2^{j_4-h}}{q_1q_2^{-1}, q_2; t q_2^d, 1, yq_2^r},
    \end{gathered}
\end{equation}
The four relations obtained from the relations of Table \ref{A2-two-term} are listed in Table \ref{A2-two-term-Higgsed} in Appendix \ref{appendix:a2-higgsed-relations-list}. These four relations are essential for deriving the bilinear relations satisfied by the tau functions of $q$-Painlev\'e VI (see the next section).
\subsection{\texorpdfstring{$A_3$ relations}{A3}}
Similar relations hold for the $A_3$ quiver partition function. We do not present the full list of these relations here and only describe their structure which is analogous to the $A_2$ case.
\begin{conjecture} \label{conjecture:A3}
Blowup relations for a (general) $A_3$ quiver partition function.
\begin{equation}
    \begin{gathered}
        f \times \Z[\mathcal{F}_6^\text{inst}]{q_1^{\theta_0}, q_1^{\theta_x}, q_1^{\sigma_0}, q_1^{\theta_t}, q_1^{\sigma_1}, q_1^{\theta_1}, q_1^{\sigma_2}, q_1^{\theta_y}, q_1^{\theta_\infty}}{q_1, q_2; x, t, 1, y} = \\
        = \sum_{\pmb{n}\in\mathbb{Z}^3+\pmb{\nu}} \C[\mathcal{C}_6]{\theta_0, \theta_x, \sigma_0, \theta_t, \sigma_1, \theta_1, \sigma_2, \theta_y, \theta_\infty}{x,t, y^{-1}}{j_0, j_1, n_0, j_2, n_1, j_3, n_2, j_4, j_5}{s,d,-r} \times \\
        \Z[\mathcal{F}_6^\text{inst}]{q_1^{\theta_0+j_0}, q_1^{\theta_x+j_1}, q_1^{\sigma_0+2n_0}, q_1^{\theta_t+j_2}, q_1^{\sigma+2n_1}, q_1^{\theta_1+j_3}, q_1^{\sigma_2+2n_2}, q_1^{\theta_y+j_4}, q_1^{\theta_\infty+j_5}}{q_1, q_1^{-1}q_2; xq_1^s, t q_1^d, 1, yq_1^r} \times \\
        \Z[\mathcal{F}_6^\text{inst}]{q_1^{\theta_0}q_2^{j_0}, q_1^{\theta_x}q_2^{j_1}, q_1^{\sigma_0}q_2^{2n_0}, q_1^{\theta_t}q_2^{j_2}, q_1^{\sigma_1}q_2^{2n_1}, q_1^{\theta_1}q_2^{j_3}, q_1^{\sigma_2}q_2^{2n_2},q_1^{\theta_y}q_2^{j_4},q_1^{\theta_\infty}q_2^{j_5}}{q_1q_2^{-1}, q_2; x q_2^s, t q_2^d, 1, yq_2^r}.
    \end{gathered}
\end{equation}
\end{conjecture}
One can introduce the Higgsing that is relevant to the $q$-Painlev\'e VI linear problem. In the $4d$ limit by AGT this 
corresponds to having two degenerate fields called $(2,1)$ and $(1,2)$ in the conformal block. With this special choice 
$\mathcal{F}_6$ gives a $q_2$ deformation of the linear problem solution constructed in \cite{JNS}. In particular, 
we recover this solution in the limit $q_2\to1$. To give an example we present some of the Higgsed relations here,
 but this list is not meant to be complete.
\begin{conjecture} \label{conjecture:A3-higgsed}
Blowup relation for the Higgsed $A_3$ quiver partition function
\begin{equation}
    \begin{gathered}
        \Z[\mathcal{F}_6^\text{inst}]{q_1^{\theta_0}, q_1q_2^2, q_1^{\theta_0}q_2^{-l}, q_1^{\theta_t}, q_1^{\sigma}, q_1^{\theta_1}, q_1^{\theta_\infty-1-h}, q_1^{-2}q_2^{-1}, q_1^{\theta_\infty-1}}{q_1, q_2; x, t, 1, y} = \\
        = \sum_{n\in\mathbb{Z}} \C[\mathcal{C}_4]{\theta_0-l \varepsilon_2/\varepsilon_1,\theta_t, \sigma, \theta_1, \theta_\infty-1-h}{t}{j_1,j_2,n,j_3,j_4}{d} \times \\
        \times \Z[\mathcal{F}_5^\text{inst}]{q_1^{\theta_0+j_1-l},q_1^{-1}q_2^2,q_1^{\theta_0+j_1}q_2^{-l}, q_1^{\theta_t+j_2}, q_1^{\sigma+2n}, q_1^{\theta_1+j_3}, q_1^{\theta_\infty-1-h+j_4}}{q_1, q_1^{-1}q_2; x q_1^s, t q_1^d, 1} \times \\
        \times \Z[\mathcal{F}_5^\text{inst}]{q_1^{\theta_0}q_2^{j_1-l}, q_1^{\theta_t}q_2^{j_2}, q_1^{\sigma}q_2^{2n}, q_1^{\theta_1}q_2^{j_3}, q_1^{\theta_\infty-1-h}q_2^{j_4},q_1^{-2}q_2,q_1^{\theta_\infty-1}q_2^{j_4-h}}{q_1q_2^{-1}, q_2; t q_2^d, 1, yq_2^r},
    \end{gathered}
\end{equation}
where
\begin{equation}
    \begin{gathered}
        (j_1,j_2,j_3,j_4,s,d,r,l,h) \in \{(1,0,0,1,1,0,1,1,1), (1,0,0,-1,1,0,1,1,-1), \\
        (-1,0,0,1,1,0,1,-1,1), (-1,0,0,-1,1,0,1,-1,-1)\}.
    \end{gathered}
\end{equation}
\end{conjecture}
\section{Tau functions and bilinear relations} \label{tau-functions-bilinear-relations}

In order to recover $q$-Painlev\'e VI bilinear relations we study the limit $q_2 \to 1$ of the Higgsed $A_2$ blowup relations. 
It produces a series of identities that express $q$-Painlev\'e VI tau functions in terms of the generating function $S$ \footnote{$S$ is a generating function of the canonical transformation that relates the initial data to a point $(y,z)$ on the trajectory of a $q$-Painlev\'e VI solution} and its derivatives (we do not use the fact that it is a generating function here, our definition of $S$ is (\ref{action-t-1-y}--\ref{action-t-y-1})). Then we take products of these identities in such a way that $S$ cancels out which leads to formulas for the solution $y(t)$ expressed via tau functions. After eliminating $y(t)$ we finally get bilinear relations.

The idea of obtaining Painlev\'e tau functions from blowup relations in the semiclassical limit was explored in \cite{JN} 
for the 4d case. We extend this program to the 5d setting. The derivation in \cite{JN} relies crucially on the Hamilton–Jacobi 
equation for the action functional, which in the 4d theory follows from the BPZ equation. A corresponding $q$-difference 
Hamilton–Jacobi equation is presently unknown, and its formulation remains an interesting open problem. Consequently, 
the argument of \cite{JN} does not admit a straightforward 5d generalization. Our approach circumvents this difficulty 
by establishing a complete set of blowup relations that closes into a system of bilinear equations for the $q$-Painlev\'e VI 
tau functions. This provides a self-contained derivation that does not rely on Hamilton–Jacobi techniques. We now turn to 
the technical details.

Following \cite{JNS} we define
\begin{equation} \label{C-tau}
    C\Bigg[\col{\theta_t}{\theta_0} \hspace{0.3em} \sigma \hspace{0.3em} \col{\theta_1}{\theta_\infty} \Bigg] = \frac{\prod_{\epsilon,\epsilon'=\pm}G_q(1+\epsilon \theta_0/2+\theta_t/2+\epsilon'\sigma/2)G_q(1+\theta_1/2+\epsilon \theta_\infty/2+\epsilon'\sigma/2)}{G_q(1+\sigma)G_q(1-\sigma)}
\end{equation}
and the tau function of the $q$-Painlev\'e VI equation
\begin{equation} \label{tau-def}
    \begin{gathered}
        \tau\Bigg[\col{\theta_t}{\theta_0} \hspace{0.3em} \sigma \hspace{0.3em} \col{\theta_1}{\theta_\infty}; s, t\Bigg] = \sum_{n \in \mathbb{Z}} s^n t^{((\sigma+2n)^2-\theta_0^2-\theta_t^2)/4} C\Bigg[\col{\theta_t}{\theta_0} \hspace{0.3em} \sigma+2n \hspace{0.3em} \col{\theta_1}{\theta_\infty}\Bigg] \times \\
        \times \Z[\mathcal{F}_4^\text{inst}]{q^{\theta_0}, q^{\theta_t}, q^{\sigma+2n}, q^{\theta_1}, q^{\theta_\infty}}{q, q^{-1}; t, 1}.
    \end{gathered}
\end{equation}
Tau function is a function of the instanton counting parameter $t$ which becomes the time in the Painlev\'e equation, the fundamental masses which become parameters of the $q$-Painlev\'e VI equation, the Coulomb modulus $\sigma$ which becomes a parameter of the $q$-Painlev\'e VI solution and another parameter of the solution $s$ which is introduced in the formula above.

We now proceed to the $q_2 \to 1$ limit of a blowup relation. We consider the limits of $C_4$ and $\mathcal{F}$ separately and then combine them to get the result. We start with the limit of $C_4$. Taking the limit of (\ref{c_n})
\begin{equation}
    \lim_{q_2\to 1}c_n(q^u) = q^{-n(n-1)(1+n+3u)/6} (q-1)^{n(n-1)/2} \frac{G_q(u+n+1)}{G_q(u+1)} \Gamma_q(u+1)^{-n}
\end{equation}
and substituting into (\ref{C_4}) we find
\begin{equation} \label{C_4 limit}
    \begin{gathered}
        \C[C_4]{\theta_0, \theta_t, \sigma, \theta_1, \theta_\infty}{t}{j_1, j_2, n, j_3, j_4}{d} \stackrel{\text{def}}{=} \lim_{q_2\to 1} \C[\mathcal{C}_4]{\theta_0, \theta_t, \sigma, \theta_1, \theta_\infty}{t}{j_1, j_2, n, j_3, j_4}{d} = \\
        = (t q^d)^{n^2}C\Bigg[\col{\theta_t+j_2}{\theta_0+j_1} \hspace{0.3em} \sigma + 2n \hspace{0.3em} \col{\theta_1+j_3}{\theta_\infty+j_4}\Bigg] C\Bigg[\col{\theta_t-1}{\theta_0} \hspace{0.3em} \sigma \hspace{0.3em} \col{\theta_1-1}{\theta_\infty}\Bigg]^{-1} \times \\
        \times \Big\{q^{(d-1)\sigma}\frac{\Gamma_q^2(1+\sigma)}{\Gamma_q^2(1-\sigma)} \prod_{\epsilon,\epsilon'=\pm} \Gamma_q^{-\epsilon'}\Big(\frac{1+\epsilon \theta_0+\theta_t+\epsilon'\sigma}{2}\Big) \Gamma_q^{-\epsilon'}\Big(\frac{1+\theta_1+\epsilon\theta_\infty+\epsilon'\sigma}{2}\Big)\Big\}^n \times \\
        \times \prod_{\epsilon,\epsilon'=\pm} \Gamma_q^{-1/2-\epsilon j_1/2-j_2/2}\Big(\frac{1+\epsilon \theta_0+\theta_t+\epsilon'\sigma}{2}\Big) \Gamma_q^{-1/2-j_3/2-\epsilon j_4/2}\Big(\frac{1+\theta_1+\epsilon\theta_\infty+\epsilon'\sigma}{2}\Big) \times \\
        \times q^{(j_2-3j_1^2j_2-j_2^3+j_3-3j_3j_4^2-j_3^3)/12 - j_1j_2\theta_0/2 + (1-j_3^2-j_4^2)\theta_1/4 + (1-j_1^2-j_2^2)\theta_t/4 - j_3j_4\theta_\infty/2} \times \\
        \times (q-1)^{(j_1^2+j_2^2+j_3^2+j_4^2)/2-1},
    \end{gathered}
\end{equation}
which means that the limit of $C_4$ reproduces the $C$-coefficient (\ref{C-tau}) needed to recover the tau function up to 
a certain modification by an exponential of $n$ and a constant which does not depend on $n$.

We now consider the limit of the partition functions. We assume that in the limit $q_2\to1$ the partition function has the 
following asymptotics
\begin{equation} \label{action-t-1-y}
    \begin{gathered}
        \Z[\mathcal{F}_5^\text{inst}]{q_1^{\theta_0}, q_1^{\theta_t}, q_1^{\sigma}, q_1^{\theta_1}, q_1^{\theta_\infty-1-h}, q_1^{-2}q_2^{-1}, q_1^{\theta_\infty-1}}{q_1, q_2; t, 1, y} = \\
        = \exp\Big(\frac{1}{\varepsilon_2}S_{t<1<y}(\theta_0,\theta_t,\sigma,\theta_1,\theta_\infty-1;q_1;t,y)+O(\varepsilon_2)\Big),
    \end{gathered}
\end{equation}
\begin{equation} \label{action-y-t-1}
    \begin{gathered}
        \Z[\mathcal{F}_5^\text{inst}]{q_1^{\theta_0}, q_1^2q_2, q_1^{\theta_0-h}, q_1^{\theta_t}, q_1^{\sigma}, q_1^{\theta_1}, q_1^{\theta_\infty-1}}{q_1, q_2; y, t, 1} = \\
        = \exp\Big(\frac{1}{\varepsilon_2}S_{y<t<1}(\theta_0,\theta_t,\sigma,\theta_1,\theta_\infty-1;q_1;t,y)+O(\varepsilon_2)\Big),
    \end{gathered}
\end{equation}
\begin{equation} \label{action-t-y-1}
    \begin{gathered}
        \Z[\mathcal{F}_5^\text{inst}]{q_1^{\theta_0}, q_1^{\theta_t}, q_1^{\sigma}, q_1^{-2}q_2^{-1}, q_1^{\sigma+h}, q_1^{\theta_1}, q_1^{\theta_\infty-1}}{q_1, q_2; t, y, 1} = \\
        = \exp\Big(\frac{1}{\varepsilon_2}S^{-}_{t<y<1}(\theta_0,\theta_t,\sigma,\theta_1,\theta_\infty-1;q_1;t,y)+O(\varepsilon_2)\Big).
    \end{gathered}
\end{equation}
When we substitute this asymptotics into a blowup relation the leading order divergency $\varepsilon_2^{-1}$ cancels and 
the subleading order gives derivatives of $S$. As for the other partition function in the lhs of a blowup relation, 
it becomes self-dual when $q_2 \to 1$.
Thus, the limit of blowup relation (\ref{blowup5-higgsed-1}) reads
\begin{equation} \label{blowup5-1-lim}
    \begin{gathered}
        f(\pmb{\theta},t,y|\pmb{j},d,r)\Big|_{q_2\to1} \times \exp\Big(q\frac{\partial S}{\partial q}-r y \frac{\partial S}{\partial y} - dt \frac{\partial S}{\partial t} \Big) \times \\
        \times \exp\Big(- \frac{1}{\log q} \Big\{(\theta_0+j_1)\frac{\partial S}{\partial \theta_0} + (\theta_t+j_2)\frac{\partial S}{\partial \theta_t} + \\
        + (\theta_1+j_3)\frac{\partial S}{\partial \theta_1} + (\theta_\infty-1-h+j_4)\frac{\partial S}{\partial \theta_\infty} + \sigma \frac{\partial S}{\partial \sigma}\Big\}\Big) = \\
        = \sum_{n \in \mathbb{Z}+\nu} \C[C_4]{\theta_0, \theta_t, \sigma, \theta_1, \theta_\infty-1-h}{t}{j_1, j_2, n, j_3, j_4}{d} \exp\Big(\frac{2n}{\log q}\frac{\partial S}{\partial \sigma}\Big) \times \\
        \times \Z[\mathcal{F}_4^\text{inst}]{q^{\theta_0+j_1}, q^{\theta_t+j_2}, q^{\sigma+2n}, q^{\theta_1+j_3}, q^{\theta_\infty-1-h+j_4}}{q, q^{-1}; t q^d, 1}.
    \end{gathered}
\end{equation}
The limit of (\ref{blowup5-higgsed-2}) is similar and the limit of (\ref{blowup5-higgsed-3}) reads
\begin{equation} \label{blowup5-2-lim}
    \begin{gathered}
        f(\pmb{\theta},t,y|\pmb{j},d,r)\Big|_{q_2\to1} \times \exp\Big(q\frac{\partial S}{\partial q}-r y \frac{\partial S}{\partial y} - dt \frac{\partial S}{\partial t}\Big) \times \\
        \times \exp\Big(- \frac{1}{\log q} \Big\{(\theta_0+j_1)\frac{\partial S}{\partial \theta_0} + (\theta_t+j_2)\frac{\partial S}{\partial \theta_t} + \\
        + (\theta_1+j_3)\frac{\partial S}{\partial \theta_1} + (\theta_\infty-1+j_4)\frac{\partial S}{\partial \theta_\infty} + \sigma \frac{\partial S}{\partial \sigma}\Big\}\Big) = \\
        = \sum_{n \in \mathbb{Z}+\nu} \C[C_5]{\theta_0, \theta_t, \sigma, -2-\frac{\varepsilon_2}{\varepsilon_1}, \sigma+h, \theta_1, \theta_\infty-1}{ty^{-1},y}{j_1, j_2, n, 2, n-\frac{h}{2}, j_3, j_4}{d-r,r} \times \\
        \times \exp\Big(\frac{2n}{\log q}\frac{\partial S}{\partial \sigma}\Big) \Z[\mathcal{F}_4^\text{inst}]{q^{\theta_0+j_1}, q^{\theta_t+j_2}, q^{\sigma+2n}, q^{\theta_1+j_3}, q^{\theta_\infty-1+j_4}}{q, q^{-1}; t q^d, 1}.
    \end{gathered}
\end{equation}

Now we can see that the lhs of these identities resembles the tau function. To make it precise we need to use (\ref{C_4 limit}) and (\ref{C_5-C_4-reduction},\ref{C_5-C_4-reduction-2}) which show that $C_5$ reduces to $C_4$ under Higgsing. Eventually we have 
\begin{equation} \label{tau1}
    \begin{gathered}
        \sum_{n \in \mathbb{Z}+\nu} \C[C_4]{\theta_0, \theta_t, \sigma, \theta_1, \theta_\infty}{t}{j_1, j_2, n, j_3, j_4}{d} \exp\Big(\frac{2n}{\log q}\frac{\partial S}{\partial \sigma}\Big) \times \\
        \times \Z[\mathcal{F}_4^\text{inst}]{q^{\theta_0+j_1}, q^{\theta_t+j_2}, q^{\sigma+2n}, q^{\theta_1+j_3}, q^{\theta_\infty+j_4}}{q, q^{-1}; t q^d, 1} = \\
        = (t q^d)^{(\theta_0+j_1)^2/4+(\theta_t+j_2)^2/4-\sigma^2/4} C\Bigg[\col{\theta_t-1}{\theta_0} \hspace{0.3em} \sigma \hspace{0.3em} \col{\theta_1-1}{\theta_\infty}\Bigg]^{-1} \times \\
        \times s_\ast^{\pm \nu} \prod_{\epsilon,\epsilon'=\pm} \Gamma_q^{-1/2-\epsilon j_1/2-j_2/2}\Big(\frac{1+\epsilon \theta_0+\theta_t+\epsilon'\sigma}{2}\Big) \Gamma_q^{-1/2-j_3/2-\epsilon j_4/2}\Big(\frac{1+\theta_1+\epsilon \theta_\infty+\epsilon'\sigma}{2}\Big) \times \\
        \times (q-1)^{(j_1^2+j_2^2+j_3^2+j_4^2)/2-1} \times \\
        \times q^{(j_2-3j_1^2j_2-j_2^3+j_3-3j_3j_4^2-j_3^3)/12 - j_1j_2\theta_0/2 + (1-j_3^2-j_4^2)\theta_1/4 + (1-j_1^2-j_2^2)\theta_t/4 - j_3j_4 \theta_\infty/2} \times \\
        \times \tau\Bigg[\col{\theta_t+j_2}{\theta_0+j_1} \hspace{1.2em} \sigma\pm2\nu \hspace{1.3em} \col{\theta_1+j_3}{\theta_\infty+j_4}; s_\ast, tq^{d}\Bigg],
    \end{gathered}
\end{equation}
where
\begin{equation} \label{s_ast-1}
    \begin{gathered}
        s_\ast = (q t)^{-\sigma}\frac{\Gamma_q^2(1+\sigma)}{\Gamma_q^2(1-\sigma)} \times \\
        \times \prod_{\epsilon,\epsilon'=\pm} \Gamma_q^{-\epsilon'}\Big(\frac{1+\epsilon \theta_0+\theta_t+\epsilon'\sigma}{2}\Big) \Gamma_q^{-\epsilon'}\Big(\frac{1+\theta_1+\epsilon \theta_\infty+\epsilon'\sigma}{2}\Big) \exp\Big(\frac{2}{\log q}\frac{\partial S}{\partial \sigma}\Big),
    \end{gathered}
\end{equation}
and
\begin{equation}
    \begin{gathered}
        \sum_{n \in \mathbb{Z}+\nu} \C[C_5]{\theta_0, \theta_t, \sigma, -2-\frac{\varepsilon_2}{\varepsilon_1}, \sigma+h, \theta_1, \theta_\infty}{ty^{-1},y}{j_1, j_2, n, 2, n-\frac{h}{2}, j_3, j_4}{d-r,r} \exp\Big(\frac{2n}{\log q}\frac{\partial S}{\partial \sigma}\Big) \times \\
        \times \Z[\mathcal{F}_4^\text{inst}]{q^{\theta_0+j_1}, q^{\theta_t+j_2}, q^{\sigma+2n}, q^{\theta_1+j_3}, q^{\theta_\infty+j_4}}{q, q^{-1}; t q^d, 1} = \\
        = (q^{\theta_1}y)^{1/4}q^{-h\sigma(1+r+j_3)/2-(2+r+j_3)/4}(q-1)\prod_{\epsilon=\pm}(1-q^{(h\sigma-\theta_1-\epsilon\theta_\infty)/2})^{(j_3+\epsilon j_4)/2} \times \\
        \times (t q^d)^{(\theta_0+j_1)^2/4+(\theta_t+j_2)^2/4-\sigma^2/4} C\Bigg[\col{\theta_t-1}{\theta_0} \hspace{0.3em} \sigma \hspace{0.3em} \col{\theta_1}{\theta_\infty}\Bigg]^{-1} \times \\
        \times s_\ast^{\pm \nu} \prod_{\epsilon,\epsilon'=\pm} \Gamma_q^{-1/2-\epsilon j_1/2-j_2/2}\Big(\frac{1+\epsilon \theta_0+\theta_t+\epsilon'\sigma}{2}\Big) \Gamma_q^{-1/2+(1-j_3)/2-\epsilon j_4/2}\Big(\frac{2+\theta_1+\epsilon \theta_\infty+\epsilon'\sigma}{2}\Big) \times \\
        \times q^{(j_2-3j_1^2j_2-j_2^3+(j_3-1)-3(j_3-1)j_4^2-(j_3-1)^3)/12 - j_1j_2\theta_0/2} \times \\
        \times q^{(1-(j_3-1)^2-j_4^2)(\theta_1+1)/4 + (1-j_1^2-j_2^2)\theta_t/4 - (j_3-1)j_4 \theta_\infty/2} \times \\
        \times (q-1)^{(j_1^2+j_2^2+(j_3-1)^2+j_4^2)/2-1} \times \\
        \times \tau\Bigg[\col{\theta_t+j_2}{\theta_0+j_1} \hspace{1.2em} \sigma\pm2\nu \hspace{1.3em} \col{\theta_1+j_3}{\theta_\infty+j_4}; s_\ast, tq^{d}\Bigg],
    \end{gathered}
\end{equation}
where

\begin{equation}
    \begin{gathered}
        s_\ast =  q^{2\sigma} \Big(\frac{q(1-q^{h\sigma})^2}{q^{\theta_1}y\prod_{\epsilon=\pm}(1-q^{(h\sigma-\theta_1-\epsilon\theta_\infty)/2})}\Big)^h \times \\
        \times (q t)^{-\sigma} \frac{\Gamma_q^2(1+\sigma)}{\Gamma_q^2(1-\sigma)} \prod_{\epsilon,\epsilon'=\pm} \Gamma_q^{-\epsilon'}\Big(\frac{1+\epsilon \theta_0+\theta_t+\epsilon'\sigma}{2}\Big) \Gamma_q^{-\epsilon'}\Big(\frac{2+\theta_1+\epsilon \theta_\infty+\epsilon'\sigma}{2}\Big) \exp\Big(\frac{2}{\log q}\frac{\partial S}{\partial \sigma}\Big).
    \end{gathered}
\end{equation}
So, when $s_\ast$ is a constant, which is achieved by taking $y=y(t)$ to be a $q$-Painlev\'e VI solution, the rhs of 
these equalities are (up to some normalization) the tau functions \cite{JNS}.
From now on we fix $h=1$ and introduce the following notation for the tau functions of $q$-Painlev\'e VI \cite{JNS}.
\begin{equation}
    \begin{alignedat}{3} \label{tau_functions}
        \tau_1 = \tau\Bigg[\col{\theta_t}{\theta_0} \hspace{1.2em} \sigma+1 \hspace{1.3em} \col{\theta_1}{\theta_\infty}; s_\ast, t\Bigg]&, \quad \tau_2 = \tau\Bigg[\col{\theta_t}{\theta_0} \hspace{0.3em} \sigma-1 \hspace{0.5em} \col{\theta_1}{\theta_\infty-2}; s_\ast, t\Bigg]&&, \\
        \tau_3 = \tau\Bigg[\col{\theta_t}{\theta_0+1} \hspace{0.3em} \sigma \hspace{0.5em} \col{\theta_1}{\theta_\infty-1}; s_\ast, t\Bigg]&, \quad \tau_4 = \tau\Bigg[\col{\theta_t}{\theta_0-1} \hspace{0.3em} \sigma \hspace{0.5em} \col{\theta_1}{\theta_\infty-1}; s_\ast, t\Bigg]&&, \\
        \tau_5 = \tau\Bigg[\col{\theta_t}{\theta_0} \hspace{0.3em} \sigma+1 \hspace{0.5em} \col{\theta_1+1}{\theta_\infty-1}; s_\ast, t\Bigg]&, \quad \tau_6 = \tau\Bigg[\col{\theta_t}{\theta_0} \hspace{0.3em} \sigma-1 \hspace{0.5em} \col{\theta_1-1}{\theta_\infty-1}; s_\ast, t\Bigg]&&, \\
        \tau_7 = \tau\Bigg[\col{\theta_t+1}{\theta_0} \hspace{0.3em} \sigma \hspace{0.6em} \col{\theta_1}{\theta_\infty-1}; s_\ast, t\Bigg]&, \quad \tau_8 = \tau\Bigg[\col{\theta_t-1}{\theta_0} \hspace{0.3em} \sigma \hspace{0.6em} \col{\theta_1}{\theta_\infty-1}; s_\ast, t\Bigg].&&
    \end{alignedat}
\end{equation}
We use the standard notation for the $q$-shifted tau functions: $\overline{\tau_i}(t)=\tau_i(q t)$, $\underline{\tau_i}(t)=\tau_i(q^{-1} t)$.
To get the bilinear relations from the blowup relations we need to take products of the $q_2 \to 1$ blowup relations in such a way that they are of total degree zero and all the $(j_1,j_2,j_3,j_4,d,r)$ sum up to $(0,0,0,0,0,0)$. 
The limit of a blowup relation (\ref{blowup5-1-lim}) can be written schematically as
\begin{equation}
    f(\pmb{\theta},t,y) \exp\Big(q\frac{\partial S}{\partial q} -\frac{1}{\log q}\frac{\partial S}{\partial \sigma}-\frac{1}{\log{q}}(\pmb{j}+\pmb{\theta}) \frac{\partial S}{\partial \pmb{\theta}} -dt\frac{\partial S}{\partial t}-ry\frac{\partial S}{\partial y}\Big) = \tau(\pmb{\theta} + \pmb{j},tq^d).
\end{equation}
Combining relations with different $(\pmb{j},d,r)$ one can exclude $S$ from these relations.
For example, if $(\pmb{j}_1,d_1,r_1)+(\pmb{j}_2,d_2,r_2)-(\pmb{j}_3,d_3,r_3)-(\pmb{j}_4,d_4,r_4)=0$, we get
\begin{equation} \label{j-cancellation-condition}
    \frac{f_1(\pmb{\theta},t,y) f_2 (\pmb{\theta},t,y)}{f_3 (\pmb{\theta},t,y) f_4(\pmb{\theta},t,y)} = \frac{\tau(\pmb{\theta}+\pmb{j}_1,tq^{d_1}) \tau(\pmb{\theta}+\pmb{j}_2,tq^{d_2})}{\tau(\pmb{\theta}+\pmb{j}_3,tq^{d_3}) \tau(\pmb{\theta}+\pmb{j}_4,tq^{d_4})}
\end{equation}

Now we apply this technique to the actual semiclassical blowup relations. Starting with the set of relations (\ref{blowup5-1-lim}) we get
\begin{equation} \label{solution}
    y = t q^{\theta_1-1} \frac{\tau_3 \tau_4}{\tau_1 \tau_2},
\end{equation}
\begin{equation} \label{solution2}
    \frac{qy-tq^{\theta_1}}{1-tq^{\theta_1}} = t q^{\theta_1-\theta_t} \frac{\underline{\tau_7} \overline{\tau_8}}{\tau_1 \tau_2},
\end{equation}
\begin{equation} \label{solution3}
    \frac{1-q^{-\theta_1+1}y}{1-tq^{\theta_t}} = \frac{\underline{\tau_5} \overline{\tau_6}}{\tau_1 \tau_2},
\end{equation}
\begin{equation} \label{solution4}
    \frac{qy-tq^{\theta_t+\theta_1}}{1-tq^{\theta_t}} = t q^{\theta_1} \frac{\tau_7 \tau_8}{\tau_1 \tau_2}.
\end{equation}
Excluding $y$ from these equations we get some of the bilinear relations on the tau functions \cite{JNS}. Namely, from 
the equations above we recover

\begin{equation} \label{bilinear1}
    \tau_1 \tau_2-q^{-\theta_t}\tau_3\tau_4+(1-tq^{\theta_t})q^{-\theta_t}\tau_7 \tau_8 = 0,
\end{equation}
\begin{equation} \label{bilinear2}
    \tau_1 \tau_2 - \tau_3\tau_4 + (1-tq^{\theta_1})q^{-\theta_t} \underline{\tau_7} \overline{\tau_8} = 0,
\end{equation}
\begin{equation} \label{bilinear3}
    \tau_1 \tau_2 - t\tau_3 \tau_4-(1-tq^{\theta_t})\underline{\tau_5}\overline{\tau_6} = 0,
\end{equation}
\begin{equation} \label{bilinear4}
    \tau_1 \tau_2 - tq^{\theta_1} \tau_3 \tau_4 - (1-tq^{\theta_1}) \tau_5 \tau_6 = 0.
\end{equation}
Note that (\ref{solution}) naturally gives a solution of $q$-Painlev\'e VI in terms of tau functions. 
Taking (\ref{blowup5-2-lim}) as a starting point it is possible to construct a solution in another region of 
convergence \footnote{Note that the instanton partition functions we use converge when the instanton counting parameters are radially ordered.}.

So far we obtained four bilinear relations satisfied by the tau functions. However, $q$-Painlev\'e VI is 
presented as a set of $8$ bilinear relations \cite{JNS}. To get the missing four bilinear relations we need to 
employ the two-term blowup relations listed in Table \ref{A2-two-term-Higgsed}. When these two-term relations 
are combined with the relations of Table \ref{Higgsed_A_2-t1y} one can exclude the partial derivatives of $S$ from 
the $q_2\to1$ limit of the two-term relations which leads to more bilinear relations on the tau functions. 
The four blowup relations of Table \ref{A2-two-term-Higgsed} give rise precisely to the four remaining bilinear 
relations on the tau functions
\begin{equation} \label{bilinear5}
    \underline{\tau_5} \tau_6+q^{\theta_0/2-\theta_t}\underline{\tau_7}\tau_8-q^{-\theta_t/2}\underline{\tau_3}\tau_4 = 0,
\end{equation}
\begin{equation} \label{bilinear6}
    \underline{\tau_5}\tau_6+q^{-\theta_0/2-\theta_t} \underline{\tau_7}\tau_8-q^{-\theta_t/2}\tau_3 \underline{\tau_4} = 0,
\end{equation}
\begin{equation} \label{bilinear7}
    \underline{\tau_5}\tau_6+q^{-1-\theta_t/2+\theta_1/2+\theta_\infty/2}t \underline{\tau_7}\tau_8 - \tau_1 \underline{\tau_2} = 0,
\end{equation}
\begin{equation} \label{bilinear8}
    \underline{\tau_5}\tau_6 + q^{-\theta_t/2+\theta_1/2-\theta_\infty/2}t \underline{\tau_7}\tau_8-\underline{\tau_1}\tau_2 = 0.
\end{equation}
Remark.
We may repeat the argument that led to (\ref{solution}--\ref{solution4}) but now looking for the linear combinations that sum up to $(0,0,0,0,0,1)$ instead of $(0,0,0,0,0,0)$. This way we obtain
\begin{equation} \label{z}
    q^{\theta_\infty/2-1}\exp\Big(-y \frac{\partial S}{\partial y}\Big) \frac{qy-tq^{\theta_1}}{qy-q^{\theta_1}} = -tq^{-\theta_t/2+\theta_1/2-1}\frac{\underline{\tau_7}\tau_8}{\underline{\tau_5}\tau_6}
\end{equation}
From \cite{JNS} we know that the rhs of (\ref{z}) is $z(t)$ - the second component of a solution of the $q$-Painlev\'e VI equation. Therefore our computation implies that
\begin{equation}
    \tilde{S}(y,\sigma,t) = S(y,\sigma,t) + \Li2(yt^{-1}q^{1-\theta_1})-\Li2(yq^{1-\theta_1}) - \log y \log(tq^{\theta_\infty/2-1})
\end{equation}
is a generating function of the canonical transformation related to the $q$-Painlev\'e VI time evolution. Namely, if $s_\ast$ is chosen according to (\ref{s_ast-1}), that is, 
\begin{equation}
    s_{\ast} \sim \exp\Big(\frac{2}{\log q}\frac{\partial \tilde S(y_\ast,\sigma,t_\ast)}{\partial \sigma}\Big)
\end{equation}
then the solution $y(t)$ determined by the given $\sigma$ and $s_\ast$ goes through the point $(y_\ast,t_\ast)$ and $z_\ast$ is given by
\begin{equation}
    \log z_\ast = -\frac{\partial{\tilde S(y_\ast,\sigma,t_\ast)}}{{\partial \log y_\ast}}
\end{equation}
at this point.
\section{Conclusion}
    In this paper we proposed several conjectures concerning blowup relations relevant to the $N_f=4$ $5$d SYM theory and 
    used them to derive the bilinear relations for the $q$-Painlev\'e VI tau functions. We considered the limits of the 
    blowup relations which involve both the $\varepsilon_1+\varepsilon_2=0$ and $\varepsilon_2\to0$ cases of the partition 
    functions in connection with the tau functions of $q$-Painlev\'e VI. In this way, we provided an explanation for the 
    appearance of gauge-theoretic quantities in the construction of tau functions of $q$-Painlev\'e VI. We used results on 
    topological string partition functions and their symmetries \cite{MPTY}, and verified that the same symmetries act on 
    the blowup relations, providing insight into the symmetries of the corresponding gauge-theory instanton partition functions.
    
    Now we point out the open problems and possible directions for further research. An interesting question to answer is 
    how the affine Weyl group action on the tau functions arises from the finite Weyl group action on the partition 
    functions / blowup relations. It would be also interesting to relate the conjectured blowup relations to the 
    earlier constructions of quantized $q$-Painlev\'e VI \cite{H}, \cite{H2} as well as more recent developments 
    reported in \cite{AHKOSS}. It is clear that blowup relations that we obtained can be employed to derive the 
    quantum $q$-Painlev\'e VI bilinear relations. So it would be interesting to find how the affine Weyl group acts 
    on the wavefunction of the quantum $q$-Painlev\'e VI, and the blowup relations that we report here could be a 
    first step to approach that problem.
\section*{Acknowledgements}
I would like to thank Jean-Emile Bourgine for his guidance throughout this research and for our regular, fruitful discussions. 
I am also grateful to Paul Zinn-Justin, Misha Bershtein, Junichi Shiraishi, Hiroaki Kanno, Koji Hasegawa, Tomoki Nosaka and 
Nick Zhu for their interest in this work and for the valuable discussions we had. I would like to thank Misha Bershtein for 
reading a draft of this paper and providing helpful feedback. Additionally, I gratefully acknowledge the support provided by 
the Australian Government Research Training Program Scholarship and the Rowden White Scholarship.
\appendix
\section{\texorpdfstring{The list of $A_1$ relations}{}} \label{appendix:a1-relations-list}
$A_1$ relations are given in the following table. A relation is described by a tuple of $(\pmb{j},d)$ in the right column. The corresponding $f$ is given in the left column.
\begin{table}[H]
    \begin{tblr}{width=1\linewidth,cells={valign=m,halign=c}, row{1,2,3,4,5,6,7,8,9,10,11,12,13,14,15,16,17,18,19}={rowsep=8pt}, colspec={|X[-1]|X[-1]|X[-1]|}}
        \hline
        $\nu$ & $f(\pmb{\theta},t|\pmb{j},d)$ & $(\pmb{j},d)$ \\
        \hline
        \SetCell[r=12]{} $0$ &   $1$ & {$(-1,0,0,-1,0),(-1,0,0,1,0),(-1,0,1,0,-1),(-1,0,1,0,0),$ \\ 
                                        $(0,-1,-1,0,1), (0,-1,0,-1,0), (0,-1,0,-1,1),$ \\ 
                                        $(0,-1,0,1,0), (0,-1,0,1,1), (0,-1,1,0,-1), (0,-1,1,0,0),$ \\ 
                                        $(0,-1,1,0,1), (0,1,1,0,-1),(1,0,0,-1,0), (1,0,0,1,0),$ \\ 
                                        $(1,0,1,0,-1), (1,0,1,0,0)$} \\
                            \hline
                            &   $\frac{1}{(1-tq_1^{\theta_t})}$ & $(1,0,-1,0,1),(-1,0,-1,0,1),(0,1,0,-1,0),(0,1,0,1,0)$ \\
                            \hline
                            &   $\frac{1}{(1-tq_1^{\theta_1})}$ & $(0,1,0,-1,-1), (1,0,-1,0,0),(0,1,0,1,-1),(-1,0,-1,0,0)$ \\
                            \hline
                            &   $\frac{1-tq_1^{-1}q_2^{-1}}{(1-tq_1^{\theta_1})}$ & $(0,1,1,0,-2)$ \\
                            \hline
                            &   $\frac{1-tq_1q_2}{(1-tq_1^{\theta_t})}$ & $(0,-1,-1,0,2)$ \\
                            \hline
                            &   $\frac{1-tq_1^{\theta_t+\theta_1+1}q_2}{(1-tq_1^{\theta_t})}$ & $(0,1,1,0,0)$ \\
                            \hline
                            &   $\frac{1-tq_1^{\theta_t+\theta_1-1}q_2^{-1}}{(1-tq_1^{\theta_1})}$ & $(0,-1,-1,0,0)$ \\
                            \hline
                            &   $\frac{1}{(1-tq_1^{\theta_t})(1-tq_1^{\theta_1})}$ & $(0,1,-1,0,0)$ \\
                            \hline
                            &   $\frac{1-tq_1^{(\theta_0+\theta_t+\theta_1+\theta_\infty+2)/2}q_2}{(1-tq_1^{\theta_t})}$ & $(1,0,0,1,1)$ \\
                            \hline
                            &   $\frac{1-tq_1^{(\theta_0+\theta_t+\theta_1-\theta_\infty+2)/2}q_2}{(1-tq_1^{\theta_t})}$ & $(1,0,0,-1,1)$ \\
                            \hline
                            &   $\frac{1-tq_1^{(-\theta_0+\theta_t+\theta_1+\theta_\infty+2)/2}q_2}{(1-tq_1^{\theta_t})}$ & $(-1,0,0,1,1)$ \\
                            \hline
                            &   $\frac{1-tq_1^{(-\theta_0+\theta_t+\theta_1-\theta_\infty+2)/2}q_2}{(1-tq_1^{\theta_t})}$ & $(-1,0,0,-1,1)$ \\
        \hline
    \end{tblr}
\end{table}
\begin{table}[H]
    \centering
    \begin{tblr}{width=1\linewidth,cells={valign=m,halign=c}, row{1,2}={rowsep=8pt}, colspec={|X[-1]|X[-1]|X[-1]|}}
        \hline
        $\nu$ & $f(\pmb{\theta},t|\pmb{j},d)$ & $(\pmb{j},d)$ \\
        \hline
        \SetCell[r=6]{} $0$ &   $\frac{1-tq_1^{(-\theta_0+\theta_t+\theta_1-\theta_\infty-2)/2}q_2^{-1}}{(1-tq_1^{\theta_1})}$ & $(1,0,0,1,-1)$ \\
                            \hline
                            &   $\frac{1-tq_1^{(-\theta_0+\theta_t+\theta_1+\theta_\infty-2)/2}q_2^{-1}}{(1-tq_1^{\theta_1})}$ & $(1,0,0,-1,-1)$ \\
                            \hline
                            &   $\frac{1-tq_1^{(\theta_0+\theta_t+\theta_1-\theta_\infty-2)/2}q_2^{-1}}{(1-tq_1^{\theta_1})}$ & $(-1,0,0,1,-1)$ \\
                            \hline
                            &   $\frac{1-tq_1^{(\theta_0+\theta_t+\theta_1+\theta_\infty-2)/2}q_2^{-1}}{(1-tq_1^{\theta_1})}$ & $(-1,0,0,-1,-1)$ \\
                            \hline
                            &   $\frac{1-tq_1^{\theta_t+1}q_2}{(1-tq_1^{\theta_t+1})(1-tq_1^{\theta_t})(1-tq_1^{\theta_t}q_2)}$ & $(0,1,-1,0,1)$ \\
                            \hline
                            &   $\frac{1-tq_1^{\theta_1-1}q_2^{-1}}{(1-tq_1^{\theta_1-1})(1-tq_1^{\theta_1})(1-tq_1^{\theta_1}q_2^{-1})}$ & $(0,1,-1,0,-1)$ \\
\hline
        \SetCell[r=4]{} $\frac{1}{2}$ &   $\mathcal{C}_{1/2}+\mathcal{C}_{-1/2}$ & {$(0, -2, 0, 0, 1), (-1, -1, 1, -1, 0), (-1, -1, 1, 1, 0),$ \\ 
                                                                                    $(-1, -1, 0, 0, 0), (-1, -1, 0, 0, 1), (-1, -1, -1, -1, 1), $ \\ 
                                                                                    $(-1, -1, -1, 1, 1), (-2, 0, 0, 0, 0), (1, -1, 1, -1, 0),$ \\ 
                                                                                    $(1, -1, 1, 1, 0), (1, -1, 0, 0, 0), (1, -1, 0, 0, 1),$ \\ 
                                                                                    $(1, -1, -1, -1, 1), (1, -1, -1, 1, 1), (0, 0, 2, 0, -1),$ \\ 
                                                                                    $(0, 0, 1, -1, -1), (0, 0, 1, -1, 0), (0, 0, 0, -2, 0),$ \\ 
                                                                                    $(0, 0, 1, 1, -1), (0, 0, 1, 1, 0), (0, 0, 0, 2, 0), $ \\
                                                                                    $(-1, 1, 1, -1, -1), (-1, 1, 1, 1, -1), (2, 0, 0, 0, 0), $ \\
                                                                                    $(1, 1, 1, -1, -1), (1, 1, 1, 1, -1), (0, 0, 0, 0, 0)$} \\
                            \hline
                            &   $\frac{(\mathcal{C}_{1/2}+\mathcal{C}_{-1/2})}{(1-tq_1^{\theta_t})}$ & {$(-1, 1, 0, 0, 0), (0, 0, -1, -1, 1), (0, 0, -1, 1, 1), (0, 0, 0, 0, 1),$ \\  $(1, 1, 0, 0, 0)$} \\
                            \hline
                            &   $\frac{(\mathcal{C}_{1/2}+\mathcal{C}_{-1/2})}{(1-tq_1^{\theta_1})}$ & {$(-1, 1, 0, 0, -1), (0, 0, -1, -1, 0), (0, 0, -1, 1, 0),$ \\ $(0, 0, 0, 0, -1), (1, 1, 0, 0, -1)$} \\
                            \hline
                            &   $\frac{(\mathcal{C}_{1/2}+\mathcal{C}_{-1/2})}{(1-tq_1^{\theta_t})(1-tq_1^{\theta_1})}$ & {$(1, 1, -1, 1, 0), (-1,1,-1,1,0), (-1,1,-1,-1,0),$ \\ $(1,1,-1,-1,0), (0,0,-2,0,1), (0,2,0,0,-1)$} \\
        \hline
    \end{tblr}
    \caption{$A_1$ relations \label{A1-table}}
\end{table}
where the shorthand notation
\begin{equation} \nonumber
    \mathcal{C}_{\nu} = \C[\mathcal{C}_4]{\theta_0,\theta_t, \sigma, \theta_1, \theta_\infty}{t}{j_1,j_2,\nu,j_3,j_4}{d}, \quad \nu=\pm1/2
\end{equation}
is used.

Note that the relation with $(j_1,j_2,j_3,j_4,d)=(0,0,0,0,0)$  has $0$ on the left due to
\begin{equation}
    \C[\mathcal{C}_4]{\theta_0,\theta_t, \sigma, \theta_1, \theta_\infty}{t}{0,0,1/2,0,0}{0} + \C[\mathcal{C}_4]{\theta_0,\theta_t, \sigma, \theta_1, \theta_\infty}{t}{0,0,-1/2,0,0}{0} = 0
\end{equation}
\section{\texorpdfstring{The list of Higgsed $A_2$ relations}{}} \label{appendix:a2-higgsed-relations-list}
The Higgsed $A_2$ relations are given in the following tables. A relation is described by a tuple of $(\pmb{j},d,r,h)$ in the right column. The corresponding $f$ is given in the left column.
\begin{table}[H]
    \centering
    \begin{tblr}{width=1\linewidth, cells={valign=m,halign=c}, row{1,2,3,4,5,6,7,8,9,10}={rowsep=8pt}, colspec={X[-1]X[-1]X[-1]}, vlines}
        \hline
        $\nu$ & $f(\pmb{\theta},t,y|\pmb{j},d,r)$ & $(\pmb{j},d,r,h)$ \\
        \hline
        \SetCell[r=3]{} $0$ & $1$ & {$(1,0,0,1,0,1,1),(0,-1,0,1,0,1,1),(-1,0,0,1,0,1,1),$ \\ $(0,-1,0,1,1,1,1), (1,0,0,-1,0,1,-1),$ \\ $(0,-1,0,-1,0,1,-1), (-1,0,0,-1,0,1,-1),$ \\ $(0,-1,0,-1,1,1,-1)$} \\
                            \hline
                            & $\frac{1-q_1^{\theta_t+\theta_1-1}ty^{-1}}{1-tq_1^{\theta_t}}$ & $(0,1,0,1,0,1,1), (0,1,0,-1,0,1,-1)$ \\
                            \hline
                            & $\frac{1-q_1^{\theta_1-1}q_2^{-1} t y^{-1}}{1-tq_1^{\theta_1}}$ & $(0,1,0,1,-1,1,1), (0,1,0,-1,-1,1,-1)$ \\
        \hline
        \SetCell[r=6]{} $\frac{1}{2}$ & $\mathcal{C}_{1/2}+\mathcal{C}_{-1/2}$ & $(0,0,0,2,0,1,1),(0,0,0,-2,0,1,-1)$ \\
                            \hline
                            & $(1-\gamma^2 y^{-1})(\mathcal{C}_{1/2}+\mathcal{C}_{-1/2})$ & $(0,0,1,1,0,2,1),(0,0,1,-1,0,2,-1)$ \\
                            \hline
                            & $(1-q_1^{\theta_1-1} y^{-1})(\mathcal{C}_{1/2}+\mathcal{C}_{-1/2})$ & $(0,0,1,1,-1,1,1),(0,0,1,-1,-1,1,-1)$ \\
                            \hline
                            & $(1-t q_1^{\theta_t})^{-1}(\mathcal{C}_{1/2}+\mathcal{C}_{-1/2})$ & $(0,0,-1,1,1,1,1),(0,0,-1,-1,1,1,-1)$ \\
                            \hline
                            & $(1-t q_1^{\theta_1})^{-1}(\mathcal{C}_{1/2}+\mathcal{C}_{-1/2})$ & $(0,0,-1,1,0,0,1),(0,0,-1,-1,0,0,-1)$ \\
                            \hline
                            & $h \gamma t^{1/4}y^{-1}\frac{q_1^{(\theta_t+3\theta_1+2\theta_\infty-4)/4}}{q_1^{\theta_\infty-1}-1}$ & $(0,0,0,0,0,1,1),(0,0,0,0,0,1,-1)$ \\
        \hline
    \end{tblr}
    \caption{Higgsed $A_2$ relations, $|\gamma^{-2}t|<1<|\gamma^2y|$\label{Higgsed_A_2-t1y}}
\end{table}
where the shorthand notation
\begin{equation} \nonumber
    \mathcal{C}_{\nu}=\C[\mathcal{C}_4]{\theta_0,\theta_t,\sigma,\theta_1,\theta_\infty-1-h}{t}{j_1,j_2,\nu,j_3,j_4}{d}, \quad \nu=\pm1/2
\end{equation}
is used.
\begin{table}[H]
    \centering
    \begin{tblr}{width=1\linewidth, cells={valign=m,halign=c}, row{1,2,3,4,5,6,7,8,9,10}={rowsep=8pt}, colspec={X[-1]X[-1]X[-1]}, vlines}
        \hline
        $\nu$ & $f(\pmb{\theta},t,y|\pmb{j},d,r)$ & $(\pmb{j},d,r,h)$ \\
        \hline
        \SetCell[r=3]{} $0$ & $1$ & {$(1,0,0,-1,0,1,1),(1,0,0,1,0,1,1),(1,0,1,0,-1,0,1),$ \\ $(1,0,1,0,0,1,1), (-1,0,0,-1,0,1,-1),$ \\ $(-1,0,0,1,0,1,-1), (-1,0,1,0,-1,0,-1),$ \\ $(-1,0,1,0,0,1,-1)$} \\
                            \hline
                            & $\frac{1-yq_1^{\theta_t+1}q_2}{1-tq_1^{\theta_t}}$ & $(1,0,-1,0,1,2,1),(-1,0,-1,0,1,2,-1)$ \\
                            \hline
                            & $\frac{1-yq_1^{\theta_t+\theta_1+1}}{1-tq_1^{\theta_1}}$ & $(1,0,-1,0,0,1,1),(-1,0,-1,0,0,1,-1)$ \\
        \hline
        \SetCell[r=6]{} $\frac{1}{2}$ & $\mathcal{C}_{1/2}+\mathcal{C}_{-1/2}$ & $(2,0,0,0,0,1,1),(-2,0,0,0,0,1,-1)$ \\
                            \hline
                            & $(1-tq_1^{\theta_t})^{-1}(\mathcal{C}_{1/2}+\mathcal{C}_{-1/2})$ & $(1,1,0,0,0,0,1),(-1,1,0,0,0,0,-1)$ \\
                            \hline
                            & $(1-tq_1^{\theta_1})^{-1}(\mathcal{C}_{1/2}+\mathcal{C}_{-1/2})$ & $(1,1,0,0,-1,0,1),(-1,1,0,0,-1,0,-1)$ \\
                            \hline
                            & $(1-\gamma^{-2}yt^{-1})(\mathcal{C}_{1/2}+\mathcal{C}_{-1/2})$ & $(1,-1,0,0,0,2,1),(-1,-1,0,0,0,2,-1)$ \\
                            \hline
                            & $(1-q_1^{\theta_t+1}yt^{-1})(\mathcal{C}_{1/2}+\mathcal{C}_{-1/2})$ & $(1,-1,0,0,1,2,1),(-1,-1,0,0,1,2,-1)$ \\
                            \hline
                            & $h \gamma^{-1} y t^{-3/4} \frac{q_1^{(2\theta_0+3\theta_t+\theta_1+2)/4}}{q_1^{\theta_0}-1}$ & $(0,0,0,0,0,1,1),(0,0,0,0,0,1,-1)$ \\
        \hline
    \end{tblr}
    \caption{Higgsed $A_2$ relations, $|\gamma^{-2}y|<|t|<|\gamma^2|$\label{Higgsed_A_2-yt1}}
\end{table}
where the shorthand notation
\begin{equation} \nonumber
    \mathcal{C}_{\nu}=\C[\mathcal{C}_4]{\theta_0-h,\theta_t,\sigma,\theta_1,\theta_\infty-1}{t}{j_1,j_2,\nu,j_3,j_4}{d}, \quad \nu=\pm1/2
\end{equation}
is used.
In the next table we omit $h$ since each tuple $(\pmb{j},d,r)$ works with both values of $h=\pm1$.
\begin{table}[H]
    \centering
    \begin{tblr}{width=1\linewidth, cells={valign=m,halign=c}, row{1,2,3,4,5,6,7,8,9,10}={rowsep=8pt}, colspec={X[-1]X[-1]X[-1]}, vlines}
        \hline
        $\nu$ & $f(\pmb{\theta},t,y|\pmb{j},d,r)$ & $(\pmb{j},d,r)$ \\
        \hline
        \SetCell[r=3]{} $0$ & $\frac{\mathcal{C}_{0,1/2}+\mathcal{C}_{0,-1/2}}{(1-yq_1^{\theta_1})(1-yq_1^{-2}q_2^{-1})}$ & {$(-1,0,0,0,-2,-1),(1,0,0,0,-2,-1),$ \\ $(0,-1,0,0,-1,-1),(0,-1,0,0,-2,-1)$} \\
                            \hline
                            & $\frac{(1-ty^{-1}q_1^{\theta_t-1})(\mathcal{C}_{0,1/2}+\mathcal{C}_{0,-1/2})}{(1-yq_1^{\theta_1})(1-yq_1^{-2}q_2^{-1})(1-q_1^{-2}q_2^{-1}tq_1^{\theta_t})}$ & $(0,1,0,0,-2,-1)$ \\
                            \hline
                            & $\frac{(1-q_1^{-1}q_2^{-1}ty^{-1})(\mathcal{C}_{0,1/2}+\mathcal{C}_{0,-1/2})}{(1-yq_1^{\theta_1})(1-q_1^{-2}q_2^{-1}tq_1^{\theta_1})(1-yq_1^{-2}q_2^{-1})}$ & $(0,1,0,0,-3,-1)$ \\
        \hline
        \SetCell[r=5]{} $\frac{1}{2}$ & $\frac{\mathcal{C}_{1/2,0}+\mathcal{C}_{-1/2,0}}{(1-yq_1^{\theta_1})(1-yq_1^{-2}q_2^{-1})}$ & $(0,0,0,1,-2,-1),(0,0,0,-1,-2,-1)$ \\
                            \hline
                            & $\frac{(1-\gamma^2 y)(\mathcal{C}_{1/2,0}+\mathcal{C}_{-1/2,0})}{1-yq_1^{\theta_1}}$ & $(0,0,1,0,-3,-2)$ \\
                            \hline
                            & $\begin{aligned} \textstyle \frac{(1-\gamma^2yq_1^{\theta_1})(\mathcal{C}_{1/2,0}+\mathcal{C}_{-1/2,0})}{(1-yq_1^{\theta_1})(1-yq_1^{\theta_1-1})(1-yq_1^{\theta_1}q_2^{-1})} \times \\ \times \textstyle \frac{1}{(1-q_1^{-2}q_2^{-1}y)(1-tq_1^{\theta_1-2}q_2^{-1})} \end{aligned}$ & $(0,0,-1,0,-2,-1)$ \\
                            \hline
                            & $\frac{(1-q_1^{-1}y)(\mathcal{C}_{1/2,0}+\mathcal{C}_{-1/2,0})}{(1-yq_1^{\theta_1})(1-q_1^{-2}y)(1-\gamma^2 y)(1-q_1^{-2}q_2^{-1}y)(1-t q_1^{\theta_t-2}q_2^{-1})}$ & $(0,0,-1,0,-1,0)$ \\
                            \hline
                            & $\frac{(1-yq_1^{\theta_1-1})(\mathcal{C}_{1/2,0}+\mathcal{C}_{-1/2,0})}{1-yq_1^{-2}q_2^{-1}}$ & $(0,0,1,0,-2,-1)$ \\
        \hline
    \end{tblr}
    \caption{Higgsed $A_2$ relations, $|\gamma^{-2}t|<|y|<|\gamma^2|$\label{Higgsed_A_2-ty1}}
\end{table}
where the shorthand notation
\begin{equation} \nonumber
    \begin{gathered}
        \mathcal{C}_{\nu_1,\nu_2}=\C[\mathcal{C}_5]{\theta_0,\theta_t,\sigma,-2-\frac{\varepsilon_2}{\varepsilon_1},\sigma+h,\theta_1,\theta_\infty-1}{ty^{-1},y}{j_1,j_2,\nu_1,2,\nu_2,j_3,j_4}{d-r,r}, \\
        \quad (\nu_1,\nu_2)\in \{(0,1/2),(0,-1/2),(1/2,0),(-1/2,0)\}
    \end{gathered}
\end{equation}
is used.

The following table contains examples of two-term Higgsed $A_2$ relations (\ref{Two-term-A2-Higgsed}).
\begin{table}[H]
    \centering
    \begin{tblr}{width=1\linewidth, cells={valign=m,halign=c}, row{1,2,3,4,5,6,7,8,9,10}={rowsep=5pt}, colspec={X[-1]X[-1]}, vlines}
        \hline
        $g_\omega$ & $\omega$ \\
        \hline
        $\Big\{\frac{1-tq_1^{\theta_1}}{1-ty^{-1}q_1^{\theta_1-1}q_2^{-1}},$  & $\{(1, 0, 0, 1, -1, 1),$ \\
        $\frac{t^{3/4}q_1^{(-2\theta_0+\theta_t+\theta_1-2\theta_\infty+1)/4}(1-tq_1^{\theta_1})(1-y^{-1}q_1^{\theta_1-1}q_2^{-1})}{q_2^{3/4}(1-ty^{-1}q_1^{\theta_1-1}q_2^{-1})}\Big\}$ & $(0, 0, -1, 1, 0, 1)\}$ \\
        \hline
        $\Big\{\frac{1-tq_1^{\theta_1}}{(1-q_1^{-\theta_0})(1-ty^{-1}q_1^{\theta_1-1}q_2^{-1})},$ & $\{(1, 0, 0, 1, -1, 1),$ \\
        $\frac{1-tq_1^{\theta_1}}{(1-q_1^{\theta_0})(1-ty^{-1}q_1^{\theta_1-1}q_2^{-1})}\Big\}$ & $(-1, 0, 0, 1, -1, 1)\}$ \\
        \hline
        $\Big\{\frac{1-tq_1^{\theta_1}}{(1-ty^{-1}q_1^{\theta_1-1}q_2^{-1})(1-tq_1^{(-\theta_0+\theta_t+\theta_1+\theta_\infty-2)/2}q_2^{-1})},$ & $\{(1, 0, 0, 1, -1, 1),$ \\
        $-\frac{t^{3/4}q_1^{(-2\theta_0+\theta_t+\theta_1-2\theta_\infty-3)/4}q_2^{-3/4}(q_1^{\theta_\infty}-q_1)(1-tq_1^{\theta_1})}{(1-tq_1^{(-\theta_0+\theta_t+\theta_1+\theta_\infty-2)/2}q_2^{-1})(1-ty^{-1}q_1^{\theta_1-1}q_2^{-1})}\Big\}$ & $(0, 0, 0, 0, -1, 1)\}$ \\
        \hline
        $\Big\{\frac{(1-tq_1^{\theta_1})}{(1-tq_1^{(-\theta_0+\theta_t+\theta_1-\theta_\infty)/2}q_2^{-1})(1-ty^{-1}q_1^{\theta_1-1}q_2^{-1})},$ & $\{(1, 0, 0, 1, -1, 1),$ \\
        $\frac{y^{-1}t^{3/4}q_1^{(-2\theta_0+\theta_t+5\theta_1+2\theta_\infty-3)/4}q_2^{-3/4}(1-tq_1^{\theta_1})}{(q_1-q_1^{\theta_\infty}q_2)(1-tq_1^{(-2\theta_0+2\theta_t+2\theta_1-2\theta_\infty)/4}q_2^{-1})(1-ty^{-1}q_1^{\theta_1-1}q_2^{-1})}\Big\}$ & $(0, 0, 0, 2, -1, 1)\}$ \\
        \hline
    \end{tblr}
    \caption{Examples of two-term Higgsed $A_2$ relations, $h=1$ \label{A2-two-term-Higgsed}}
\end{table}
\section{Coefficients} \label{appendix:coefficients}
We use the following functions as the building blocks of the coefficients in the blowup relations.
\begin{equation} \label{c_n}
    c_n(v) = \prod_{\substack{i,j\geq 0 \\ i+j\leq n-2}}(1-q_1^{-i-1}q_2^{-j-1}v^{-1}) \prod_{\substack{i,j\geq0 \\ i+j\leq-n-1}}(1-q_1^i q_2^j v^{-1})
\end{equation}
\begin{equation}
    g_{n}(v) = (-v)^{n(n-1)/2}\gamma^{-n(n-1)(n+1)/3}
\end{equation}
These functions enjoy the following symmetries
\begin{equation}
    g_n(v^{-1}) c_n(v^{-1}) = (-v)^{-n(n-1)/2} \gamma^{-n(n-1)(n+1)/3} c_n(v^{-1}) = c_{-n+1}(\gamma^2 v) = c_{n}(v)\Big|_{q_1\to q_1^{-1},q_2\to q_2^{-1}}
\end{equation}
\begin{equation}
    \begin{gathered}
        c_{n+2}(\gamma^2 v) = \frac{\gamma^{n(n+1)}(v-1)}{v^{2n+1}} \frac{\prod_{i=1}^n(1-vq_1^i)(1-vq_2^i)}{\prod_{i=1}^{-n}(1-vq_1^{-i+1})(1-vq_2^{-i+1})}c_n(v) = \\
        = \frac{\gamma^{n(n+1)}(v,q_1)_{n+1} (v,q_2)_{n+1}}{v^{2n+1}(v-1)} c_n(v)
    \end{gathered}
\end{equation}
\begin{equation}
    \frac{c_{-n+1}(v^{-1}q_1^{-1})}{c_{-n}(v^{-1})} = (v,q_2)_n^{-1}
\end{equation}
and
\begin{equation}
    c_0 = c_1 = 1
\end{equation}
We also note the following asymptotics for $c_n$
\begin{equation}
    c_n(v) \sim v^{n(n-1)/2}, v \to 0
\end{equation}
The coefficients that appear in the blowup relations written on the instanton parts of the partition function are
\begin{equation} \label{C_4}
    \begin{gathered}
        \C[\mathcal{C}_4]{\theta_0, \theta_t, \sigma, \theta_1, \theta_\infty}{t}{j_1, j_2, n, j_3, j_4}{d} = \\
        = q_1^{\sigma n(1+j_2+j_3+d)} \Big(\frac{t q_1^{\theta_1+\theta_t}}{\gamma^{2(j_2+j_3+d)}}\Big)^{n^2} \times \\
        \times \frac{\prod_{\epsilon,\epsilon'=\pm}c_{n\epsilon'+(1+\epsilon j_1+j_2)/2}\Big(\gamma q_1^{(\epsilon\theta_0+\theta_t+\epsilon'\sigma)/2}\Big)c_{n\epsilon' + (1+j_3+\epsilon j_4)/2}\Big(\gamma q_1^{(\theta_1+\epsilon\theta_\infty+\epsilon'\sigma)/2}\Big)}{c_{2n}(q_1^\sigma)c_{-2n}(q_1^{-\sigma})}
    \end{gathered}
\end{equation}
Blowup relations can be rewritten in terms of the Weyl group invariant partition function $\mathcal{F}$.
Using the following property of the double Pochhammer symbol
\begin{equation}
    \begin{gathered}
        \frac{(u;q_1,q_2)_\infty}{(u q_1^{n};q_1,q_1^{-1} q_2)_\infty (u q_2^{n};q_1 q_2^{-1},q_2)_\infty} = c_{-n}(u^{-1}), \\
        \frac{\Gamma(u;q_1,q_2)}{\Gamma(u q_1^{n};q_1,q_1^{-1} q_2) \Gamma(u q_2^{n};q_1 q_2^{-1},q_2)} = g_{-n}(u^{-1})
    \end{gathered}
\end{equation}
and, more generally,
\begin{equation}
    \begin{gathered}
        \frac{((q_1 q_2)^{-k/2} u;q_1,q_2)_\infty}{(q_2^{-k/2}uq_1^n;q_1,q_1^{-1} q_2)_\infty (q_1^{-k/2} u q_2^n;q_1 q_2^{-1},q_2)_\infty} = \\
        = \frac{(\gamma^k u;q_1,q_2)_\infty}{(\gamma^k u q_1^{n+k/2};q_1,q_1^{-1}q_2)_\infty (\gamma^k uq_2^{n+k/2};q_1 q_2^{-1},q_2)_\infty} = c_{-n-k/2}(\gamma^{-k} u^{-1}),
    \end{gathered}
\end{equation}
we get
\begin{equation}
    \begin{gathered}
        \Z[\mathcal{F}^{\text{1-loop}}_4]{q_1^{\theta_0}, q_1^{\theta_t}, q_1^{\sigma}, q_1^{\theta_1}, q_1^{\theta_\infty}}{q_1, q_2} \times \\
        \times \Z[\mathcal{F}^{\text{1-loop}}_4]{q_1^{\theta_0+j_1}, q_1^{\theta_t+j_2}, q_1^{\sigma+2n}, q_1^{\theta_1+j_3}, q_1^{\theta_\infty+j_4}}{q_1, q_1^{-1} q_2}^{-1} \times \\
        \times \Z[\mathcal{F}^{\text{1-loop}}_4]{q_1^{\theta_0}q_2^{j_1}, q_1^{\theta_t}q_2^{j_2}, q_1^{\sigma}q_2^{2n}, q_1^{\theta_1}q_2^{j_3}, q_1^{\theta_\infty}q_2^{j_4}}{q_1 q_2^{-1}, q_2}^{-1} = \\
        = \frac{g_{2n}(q_1^\sigma)}{\prod_{\epsilon = \pm} g_{n+(-\epsilon j_1+j_2+1)/2}(\gamma q_1^{(-\epsilon \theta_0+\theta_t+\sigma)/2}) g_{n+(j_3-\epsilon j_4+1)/2}(\gamma q_1^{(\theta_1-\epsilon \theta_\infty+\sigma)/2})} \times \\
        \times q_1^{n \sigma(1+j_2+j_3+d)} \Big(\frac{t q_1^{\theta_t+\theta_1}}{\gamma^{2(j_2+j_3+d)}}\Big)^{n^2} \C[\mathcal{C}_4]{\theta_0, \theta_t, \sigma, \theta_1, \theta_\infty}{t}{j_1, j_2, n, j_3, j_4}{d}^{-1},
    \end{gathered}
\end{equation}
\begin{equation}
    \begin{gathered}
        \frac{\mathcal{F}_{4}^{\text{extra}}(q_1^{\theta_t},q_1^{\theta_1},t;q_1,q_2)}{\mathcal{F}_{4}^{\text{extra}}(q_1^{\theta_t+j_2},q_1^{\theta_1+j_3},tq_1^d;q_1,q_1^{-1}q_2)\mathcal{F}_{4}^{\text{extra}}(q_1^{\theta_t}q_2^{j_2},q_1^{\theta_1}q_2^{j_3},tq_2^d;q_1 q_2^{-1},q_2)} = \\
        = c_{-j_2-d}(t^{-1}q_1^{-\theta_t})^{-1} c_{1-j_3-d}(\gamma^2 t^{-1}q_1^{-\theta_1})^{-1}.
    \end{gathered}
\end{equation}
Therefore, it is convenient to define
\begin{equation} \label{A_4}
    \begin{gathered}
        \C[\mathcal{A}_4]{\theta_0,\theta_t,\sigma,\theta_1,\theta_\infty}{t}{j_1,j_2,n,j_3,j_4}{d} =
        q_1^{n \sigma(1+j_2+j_3+d)} \Big(\frac{t q_1^{\theta_t+\theta_1}}{\gamma^{2(j_2+j_3+d)}}\Big)^{n^2} c_{-j_2-d}(t^{-1}q_1^{-\theta_t})^{-1} \times \\
        \times c_{1-j_3-d}(\gamma^2 t^{-1}q_1^{-\theta_1})^{-1} \frac{g_{2n}(q_1^\sigma)}{\prod_{\epsilon = \pm} g_{n+(\epsilon j_1+j_2+1)/2}(\gamma q_1^{(\epsilon \theta_0+\theta_t+\sigma)/2}) g_{n+(j_3+\epsilon j_4+1)/2}(\gamma q_1^{(\theta_1+\epsilon \theta_\infty+\sigma)/2})}.
    \end{gathered}
\end{equation}
These coefficients appear in the blowup relations that the Weyl group - invariant partition functions obey
\begin{equation}
    \begin{gathered} \label{blowup4_inv}
        f(\pmb{\theta},t|\pmb{j},d)\times\Z[\mathcal{F}_4]{q_1^{\theta_0}, q_1^{\theta_t}, q_1^{\sigma}, q_1^{\theta_1}, q_1^{\theta_\infty}}{q_1, q_2; t, 1} = \sum_{n\in\mathbb{Z}+\nu} \C[\mathcal{A}_4]{\theta_0, \theta_t, \sigma, \theta_1, \theta_\infty}{t}{j_1, j_2, n, j_3, j_4}{d} \times \\
        \times \Z[\mathcal{F}_4]{q_1^{\theta_0+j_1}, q_1^{\theta_t+j_2}, q_1^{\sigma+2n}, q_1^{\theta_1+j_3}, q_1^{\theta_\infty+j_4}}{q_1, q_1^{-1}q_2; t q_1^d, 1} \times \\
        \times \Z[\mathcal{F}_4]{q_1^{\theta_0}q_2^{j_1}, q_1^{\theta_t}q_2^{j_2}, q_1^{\sigma}q_2^{2n}, q_1^{\theta_1}q_2^{j_3}, q_1^{\theta_\infty}q_2^{j_4}}{q_1q_2^{-1}, q_2; t q_2^d, 1}.
    \end{gathered}
\end{equation}
This form of blowup relations allows us to compute $f$ factors appearing in the lhs of them. Applying the symmetry (\ref{A1:S1}--\ref{A1:S5}) to a blowup relation, one gets (\ref{A1:f_action}).

In the $A_2$ quiver case the corresponding coefficients are given by
\begin{equation} \label{C_5}
    \begin{gathered}
        \C[\mathcal{C}_5]{\theta_0, \theta_t, \sigma_1, \theta_1, \sigma_2, \theta_y, \theta_\infty}{t_1, t_2}{j_1, j_2, n_1, j_3, n_2, j_4, j_5}{d_1,d_2} = \\
        = q_1^{\sigma_1 n_1(1+j_2+j_3+d_1)+\sigma_2 n_2(1+j_3+j_4+d_2)} \Big(\frac{t_1 q_1^{\theta_t+\theta_1}}{\gamma^{2(j_2+j_3+d_1)}}\Big)^{n_1^2} \Big(\frac{t_2 q_1^{\theta_1+\theta_y}}{\gamma^{2(j_3+j_4+d_2)}}\Big)^{n_2^2} \times \\
        \times \frac{\prod_{\epsilon,\epsilon'=\pm}c_{n_1\epsilon'+(1+\epsilon j_1+j_2)/2}\Big(\gamma q_1^{(\epsilon\theta_0+\theta_t+\epsilon'\sigma_1)/2}\Big)c_{n_1\epsilon'+n_2\epsilon+(1+j_3)/2}\Big(\gamma q_1^{(\sigma_1\epsilon'+\theta_1+\sigma_2\epsilon)/2}\Big)}{c_{2n_1}(q_1^{\sigma_1})c_{-2n_1}(q_1^{-\sigma_1})c_{2n_2}(q_1^{\sigma_2})c_{-2n_2}(q_1^{-\sigma_2})} \times \\
        \times \prod_{\epsilon,\epsilon'=\pm} c_{n_2\epsilon + (1+j_4+\epsilon' j_5)/2}\Big(\gamma q_1^{(\epsilon\sigma_2+\theta_y+\epsilon'\theta_\infty)/2}\Big)
    \end{gathered}
\end{equation}
and
\begin{equation} \label{A_5}
    \begin{gathered}
        \C[\mathcal{A}_5]{\theta_0, \theta_t, \sigma_1, \theta_1, \sigma_2, \theta_y, \theta_\infty}{t_1, t_2}{j_1, j_2, n_1, j_3, n_2, j_4, j_5}{d_1,d_2} = \\
        = q_1^{n_1\sigma_1(1+d_1+j_2+j_3)+n_2\sigma_2(1+d_2+j_3+j_4)} \Big(\frac{t_1q_1^{\theta_t+\theta_1}}{\gamma^{2(d_1+j_2+j_3)}}\Big)^{n_1^2} \Big(\frac{t_2q_1^{\theta_1+\theta_y}}{\gamma^{2(d_2+j_3+j_4)}}\Big)^{n_2^2} \times \\
        \times \frac{c_{1/2-d_1-j_3/2-n_1+n_2}(\gamma t_1^{-1} q_1^{(-\sigma_1-\theta_1+\sigma_2)/2}) c_{1/2-d_1-j_2-n_1-j_3/2+n_2}(\gamma t_1^{-1} q_1^{-\theta_t-(\sigma_1+\theta_1-\sigma_2)/2})}{c_{-d_1-j_2}(t_1^{-1} q_1^{-\theta_t}) c_{1-d_1-j_3}(\gamma^{2} t_1^{-1} q_1^{-\theta_1}) c_{-d_1-d_2-j_2-j_3}(t_1^{-1}t_2^{-1} q_1^{-\theta_1-\theta_t})} \times \\
        \times \frac{c_{1/2-d_2+n_1-j_3/2-n_2}(\gamma t_2^{-1} q_1^{(\sigma_1-\theta_1-\sigma_2)/2}) c_{1/2-d_2-j_4+n_1-j_3/2-n_2}(\gamma t_2^{-1} q_1^{-\theta_y+(\sigma_1-\theta_1-\sigma_2)/2})}{c_{1-d_1-d_2-j_3-j_4}(\gamma^{2} t_1^{-1}t_2^{-1} q_1^{-\theta_1-\theta_y}) c_{-d_2-j_3}(t_2^{-1} q_1^{-\theta_1}) c_{1-d_2-j_4}(\gamma^{2} t_2^{-1} q_1^{-\theta_y})} \times \\
        \times c_{1/2-n_1+j_3/2+n_2}(\gamma q_1^{(-\sigma_1+\theta_1+\sigma_2)/2}) c_{1/2+n_1+j_3/2-n_2}(\gamma q_1^{(-\sigma_2+\theta_1+\sigma_1)/2}) \times \\
        \times \frac{g_{2n_1}(q_1^{\sigma_1})g_{2n_2}(q_1^{\sigma_2})}{g_{n_1+n_2+(1+j_3)/2}(\gamma q_1^{(\sigma_1+\theta_1+\sigma_2)/2})} \times \\
        \times (\prod_{\epsilon=\pm}g_{n_1+(1+\epsilon j_1+j_2)/2}(\gamma q_1^{(\epsilon \theta_0+\theta_t+\sigma_1)/2})g_{n_2+(1+j_4+\epsilon j_5)/2}(\gamma q_1^{(\sigma_2+\theta_y+\epsilon \theta_\infty)/2}))^{-1}.
    \end{gathered}
\end{equation}
Similarly to $\mathcal{A}_4$, the formula for $\mathcal{A}_5$ follows from the Weyl group - invariant partition function (\ref{F-5-full}).

We also use shorthand notation
\begin{equation}
    \mathcal{C}_{\nu_1,\nu_2} = \C[\mathcal{C}_5]{\theta_0, \theta_t, \sigma_1, \theta_1, \sigma_2, \theta_y, \theta_\infty}{t_1, t_2}{j_1, j_2, \nu_1, j_3, \nu_2, j_4, j_5}{d_1,d_2}
\end{equation}
when other parameters are clear from the context.

Now we proceed to the formulas that describe $\mathcal{C}_5 \to \mathcal{C}_4$ reduction under Higgsing.
\begin{equation} \label{C_5-C_4-reduction}
    \begin{gathered}
        \C[\mathcal{C}_5]{\theta_0, \theta_t, \sigma, \theta_1, \theta_\infty-1-h, -2-\frac{\varepsilon_2}{\varepsilon_1} \quad, \theta_\infty-1}{t, y^{-1}}{j_1, j_2, n_1, j_3, n_2, 2, j_4-h}{d,-r} = \\
        = \delta_{j_4,2n_2}(1-q_1) q_1^{j_4(3+j_3-r)(\theta_\infty-1-h)/2} (y^{-1}(q_1q_2)^{j_3+2-r}q_1^{\theta_1}(q_1^{-2}q_2^{-1}))^{j_4^2/4} \times \\
        \times \prod_{\epsilon=\pm} \frac{c_{3/2+\epsilon(h/2-j_4)}(q_1^{\epsilon(1+h/2)-3/2}q_2^{-1}q_1^{-\epsilon \theta_\infty})}{c_{\epsilon j_4}(q_1^{\epsilon(\theta_\infty-1-h)})}\times \\
        \times \C[\mathcal{C}_4]{\theta_0, \theta_t, \sigma, \theta_1, \theta_\infty-1-h}{t}{j_1, j_2, n_1, j_3, j_4}{d},
    \end{gathered}
\end{equation}
\begin{equation} \label{C_5-C_4-reduction-2}
    \begin{gathered}
        \C[\mathcal{C}_5]{\theta_0, \theta_t, \sigma, -2-\frac{\varepsilon_2}{\varepsilon_1} \quad, \sigma+h, \theta_1, \theta_\infty}{t_1, t_2}{j_1, j_2, n_1, 2, n_2, j_3, j_4}{d_1,d_2} = \\
        = q_1^{(-2-d_2-j_3)/4+hn_1}q_2^{(1+d_2+j_3)/4-h(d_2+j_3)n_1} \times \\ 
        \times \frac{\delta_{n_1,n_2+h/2} (q_1^{\theta_1}t_2)^{1/4-hn_1}(q_1^{\sigma})^{-h(1+d_2+j_3)/2+2n_1}(q_1-1)(q_1^{h \sigma};q_2)_{2h n_1}}{\prod_{\epsilon=\pm}(\gamma^{-1} q_1^{(h\sigma-\theta_1-1-\epsilon \theta_\infty)/2};q_2)_{hn_1-(1+j_3-1+\epsilon j_4)/2}} \times \\
        \times \C[\mathcal{C}_4]{\theta_0, \theta_t, \sigma, \theta_1+1, \theta_\infty}{q_2 t_1 t_2}{j_1, j_2, n_1, j_3-1, j_4}{d_1+d_2}.
    \end{gathered}
\end{equation}
Taking the limit $q_2 \to 1$ and using the formula for $C_4$ (\ref{C_4 limit}) we get
\begin{equation}
    \begin{gathered}
        \lim_{q_2 \to 1}\C[\mathcal{C}_5]{\theta_0, \theta_t, \sigma, -2-\frac{\varepsilon_2}{\varepsilon_1} \quad, \sigma+h, \theta_1, \theta_\infty}{t_1, t_2}{j_1, j_2, n_1, 2, n_2, j_3, j_4}{d_1,d_2} = \\
        = \delta_{n_1,n_2+h/2} (q^{d_1+d_2} t_1 t_2)^{n_1^2} C\Bigg[\col{\theta_t+j_2}{\theta_0+j_1} \hspace{0.3em} \sigma + 2n_1 \hspace{0.3em} \col{\theta_1+j_3}{\theta_\infty+j_4}\Bigg] C\Bigg[\col{\theta_t-1}{\theta_0} \hspace{0.3em} \sigma \hspace{0.3em} \col{\theta_1}{\theta_\infty}\Bigg]^{-1} \times \\
        \times \Big\{\Big(\frac{q^{1-\theta_1}t_2^{-1}(1-q^{h\sigma})^2}{\prod_{\epsilon=\pm}(1-q^{(h\sigma-\theta_1-\epsilon\theta_\infty)/2})}\Big)^h q^{(d_1+d_2+1)\sigma} \times \\
        \times \frac{\Gamma_q^2(1+\sigma)}{\Gamma_q^2(1-\sigma)} \prod_{\epsilon,\epsilon'=\pm} \Gamma_q^{-\epsilon'}\Big(\frac{1+\epsilon \theta_0+\theta_t+\epsilon'\sigma}{2}\Big) \Gamma_q^{-\epsilon'}\Big(\frac{2+\theta_1+\epsilon\theta_\infty+\epsilon'\sigma}{2}\Big)\Big\}^{n_1} \times \\
        \times \prod_{\epsilon,\epsilon'=\pm} \Gamma_q^{-1/2-\epsilon j_1/2-j_2/2}\Big(\frac{1+\epsilon \theta_0+\theta_t+\epsilon'\sigma}{2}\Big) \Gamma_q^{-j_3/2-\epsilon j_4/2}\Big(\frac{2+\theta_1+\epsilon\theta_\infty+\epsilon'\sigma}{2}\Big) \times \\
        \times q^{(j_2-3j_1^2j_2-j_2^3+(j_3-1)-3(j_3-1)j_4^2-(j_3-1)^3)/12 - j_1j_2\theta_0/2} \times \\
        \times q^{(1-(j_3-1)^2-j_4^2)(\theta_1+1)/4 + (1-j_1^2-j_2^2)\theta_t/4 - (j_3-1)j_4\theta_\infty/2} \times \\
        \times (q^{\theta_1}t_2)^{1/4}q^{-h\sigma(1+d_2+j_3)/2-(2+d_2+j_3)/4}(q-1)^{(j_1^2+j_2^2+(j_3-1)^2+j_4^2)/2} \times \\
        \times \prod_{\epsilon = \pm}(1-q^{(h\sigma-\theta_1-\epsilon \theta_\infty)/2})^{(j_3+\epsilon j_4)/2}.
    \end{gathered}
\end{equation}
\newpage
\bibliographystyle{JHEP}
\bibliography{references}
\nocite{*}
\end{document}